\documentclass[12pt]{iopart}
\usepackage{iopams}
\usepackage{graphicx}  
\usepackage{dcolumn}   
\usepackage{bm}        
\usepackage{amssymb}   
\usepackage[latin1]{inputenc}
\usepackage{amsfonts}
\expandafter\let\csname equation*\endcsname\relax
\expandafter\let\csname endequation*\endcsname\relax
\usepackage{amsmath}
\usepackage{amsthm}
\usepackage{xcolor}
\usepackage{fontenc}
\usepackage{textcomp}
\usepackage{epstopdf}
\usepackage{hyperref}

\begin{document}

\title{Critical and quasicritical behavior in a three-species dynamical model of semi-directed percolation}

\author{C. K. Jasna}
\ead{jasnack@cusat.ac.in, jasnack00@gmail.com}
\address{Department of Physics, Cochin University of Science and Technology, Cochin 682022, India.}
\author{V. Sasidevan}
\ead{sasidevan@cusat.ac.in, sasidevan@gmail.com}
\address{Department of Physics, Cochin University of Science and Technology, Cochin 682022, India.}

\date{\today}

\begin{abstract}
We investigate a one-dimensional three-species dynamical model whose dynamics naturally generate the semi-directed percolation cluster in time and show a non-equilibrium absorbing state phase transition from an active to inactive state. The critical threshold and exponents associated with the dynamic process are determined using Monte Carlo simulations. Critical behavior observed shows that the model belongs to the directed percolation (DP) universality class. Further, we consider the effect of spontaneous activity generation in the dynamical model. While, as expected, this destroys the usual critical behaviour, we find that the dynamic susceptibility shows a maximum at a specific value of the control parameter, indicating a quasi-critical behaviour, similar to the findings in the case of DP models and DP-inspired models of neuronal activity with spontaneous activity generation. Interestingly, in the presence of spontaneous activity, we find that spatial and temporal correlations exhibit power-law decays at a value of the control parameter different from the pseudo-threshold corresponding to the peak of the dynamic susceptibility, indicating that there are two pseudo-thresholds in such a case, one where the response function is maximum and another where the spatial and temporal correlations showscale-freee behaviour. 
\end{abstract}

\maketitle

\noindent{\it Keywords\/}:
Semi-directed percolation, absorbing state phase transition, spontaneous activity, quasicritical behavior.

\section{Introduction}

Phase transitions in percolation models have been of long-standing interest in statistical physics and have been analyzed extensively both from theoretical and application points of view in diverse fields~\cite{Saberi2015,Sahimi2021,Hunt2014, Ziff2021,Panda2023,Isichenko1992}. Numerous variants of the percolation model have been introduced and studied in the past, and one of the variants that still poses many interesting questions is the model of directed percolation~\cite{Henkel, Hinrichsen2000}. Unlike the standard or isotropic percolation model, the directed model has an inbuilt preferred direction that can be associated with the evolution of a dynamic process. Usually, a temporal coordinate is introduced to denote the preferred direction, which in the two-dimensional case leads to a space-like plus a time-like dimension, resulting in the so-called $1+1$-dimensional directed percolation model~\cite{Henkel, Hinrichsen2000}. Similar to the isotropic percolation model, the phase transition in the directed model is marked by the appearance of an infinite cluster that extends over the system, but with a preferred direction~\cite{Hinrichsen2000,Dhar1981}. While the directed cluster can be constructed from an isotropic one, a convenient and useful way to generate and study the former is by defining a suitable dynamic process, as is usually done when defining the directed percolation problem~\cite{Hinrichsen2000}.  

There exist many similarities and differences between the isotropic and directed models of percolation. The order parameter for both cases is the probability that a randomly chosen site belongs to/generates an infinite cluster of occupied or active sites. However, the nature of the phase transition and associated critical properties in the two models are different. In contrast to the isotropic percolation (IP) model, directed percolation (DP) models are inherently non-equilibrium in nature due to the presence of absorbing state configurations in the dynamic process from which the system cannot escape once entered~\cite{Henkel}. By and large, the IP and the DP universality classes form two major groups, the former characterizing many purely geometric or static processes and the latter characterizing many dynamic processes leading to absorbing states~\cite{Hinrichsen2000,Hinrichsen2009,Janssen1981,Grassberger1982}.

Delineating the relationship between IP and DP has been a subject of significant interest. Earlier, a few studies have investigated the crossover from isotropic to directed critical behaviour in suitably defined percolation models, aiming to identify the mechanisms and parameters that could control this transition~\cite{Hinrichsen2000}. In these models, it is often observed that the critical behavior shows a shift from one universality class to the other when a parameter of the system is varied. For example, in Ref.~\cite{Frojdh1997}, the authors introduced a fugacity parameter which, when tuned, changes the percolation system from isotropic to directed. In another study~\cite{Zhou2012}, a biased directed percolation model was introduced with a probability parameter that causes a crossover from isotropic to directed percolation at a specific value of the parameter.

An interesting variant in this context is the semi-directed percolation (SDP) model introduced by Martin and Vannimenus~\cite{HOMartin1985}, which occupies an intermediate position between isotropic and fully directed percolation models and is the focus of our present work. The model exhibits directionality along one of the dimensions while remaining isotropic in others. Its critical thresholds and correlation length exponent were evaluated in Ref.\cite{HOMartin1985,Knezevic2016} using transfer matrix and phenomenological renormalization techniques.

In this work, our first contribution is to formulate the SDP problem as a one-dimensional dynamical model. In particular, we define a one-dimensional three-species model whose dynamics naturally generate the semi-directed percolation cluster in time and show an absorbing state phase transition from an active to inactive state. At first, using Monte-Carlo simulations, we determine the threshold and critical exponents, verifying numerically the earlier theoretical predictions~\cite{HOMartin1985,Knezevic2016} and confirming that the dynamical SDP model belongs to the fully directed percolation (DP) universality class. Beyond this, the proposed dynamic framework allows us to consider the SDP problem as involving an active-absorbing phase transition happening in time and study the critical exponents associated with the dynamic process. Further, aided by the dynamical framework, we consider the effect of spontaneous activity generation in SDP. Dynamical models similar to DP with spontaneous activity generation have shown promise in modeling biological systems, such as cardiac and neuronal activity~\cite{Rabinovitch2021, Molla2024} and in studies of forest fire dynamics~\cite{Drossel1993,Jensen2021}. In such models, spontaneous activity generation corresponds to activation of neural cells or burning of trees by an external mechanism (such as lightning in the case of a forest fire), which then can combine with the internally generated activity, aiding its spreading through the system. 

In the dynamical SDP model we consider,  while the spontaneous activity generation expectedly destroys the usual critical behaviour, we find that the dynamic susceptibility still shows a maximum as a function of the control parameter defining a pseudo-threshold~\cite{Fosque2021}. Similar findings have been reported in the case of DP models in the presence of an external field~\cite{Henkel,Hinrichsen2000,Lubeck2002,Lubeck2004,Lubeck2004b,Lubeck2005} and  DP-like models of neuronal dynamics with spontaneous activity~\cite{Korchinski2021,WilliamsGarcia2014,Fosque2021,Fosque2022,beggs2022,Tian2022}. We obtain the non-equilibrium Widom line corresponding to the maxima of the dynamic susceptibility for different strengths of spontaneous activity generation. Curiously, we find that spatial and temporal correlations exhibit power-law decays at a value of the control parameter different from the pseudo-threshold corresponding to the peak of the dynamic susceptibility. This indicates that in the presence of spontaneous activity, there are two pseudo-thresholds, one where the response function is maximum and another where the spatial and temporal correlations show scale free behaviour.

The paper is structured as follows. In Sec.~\ref{section1}, we precisely define the three-species model and the dynamic process. In Sec.~\ref{section2}, we obtain estimates of thresholds and critical exponents of SDP model using Monte Carlo simulations. The effect of introducing spontaneous activity is considered in Sec.~\ref{subsection2}. Spatial and temporal correlations, as well as dynamical susceptibility, are obtained for different strengths of spontaneous activity. The associated critical thresholds and exponents are determined. Finally, we conclude in Sec.~\ref{section3}.

\section{Model definition}
\label{section1}
The dynamical version of the Semi-Directed Percolation (SDP) problem consists of a one-dimensional lattice in which each site can be in either of the three states denoted by $0$, $1$, and $2$.  Here, $0$ denotes the inactive {\em and} immune state, and $2$ represents the active state. State $1$ denotes a susceptible state for the spread of activity. The dynamical evolution involves the spontaneous change of states of sites and the spreading of activity through clusters of susceptible sites at each time step.  These two processes are: P1) Spontaneous change of an active site to an immune one ($2 \rightarrow 0$) and a change between susceptible and immune states ($1 \longleftrightarrow 0$), and P2) Spreading of activity through each maximal contiguous set of susceptible (1s) sites due to contact with an active site ($1 \rightarrow 2$ ). Each of the state changes in the former occurs independently and probabilistically, while the latter process occurs with probability one. Spontaneous generation of activity is considered by including the possibility of each maximal cluster of 1s turning into 2 with a probability $\epsilon$ independently, via P1. Starting from an initial configuration at time $t=0$, the specific dynamical rules for the two processes are: 

P1) \\
\begin{equation}
    0 \rightarrow \begin{cases}
        1 &  \textrm{with prob p}\\
        0 & \textrm{with prob (1-p)}
    \end{cases}
    \label{a1}
    \end{equation}

     \begin{equation}
     \begin{aligned}
\textrm{Cluster of ~ 1s} \rightarrow 2~~~  
    \textrm{with probability $\epsilon$ } \\ 
    \textrm{or follow}
    ~\textrm{Eq.} ~\ref{a3}~
    \textrm{with prob} ~ (1-\epsilon).
     \label{a2}
    \end{aligned}
    \end{equation}
\begin{equation}
    1 \rightarrow \begin{cases}
        1 &  \textrm{with prob p}\\
        0 & \textrm{with prob (1-p)}
        \end{cases}
         \label{a3}
    \end{equation}

      \begin{equation}
    2 \rightarrow \begin{cases}
        2 &  \textrm{with prob p}\\
        0 & \textrm{with prob (1-p)}
    \end{cases}
     \label{a4}
    \end{equation}

\begin{figure*}[!ht]
    \centering
    \includegraphics[scale=0.34]{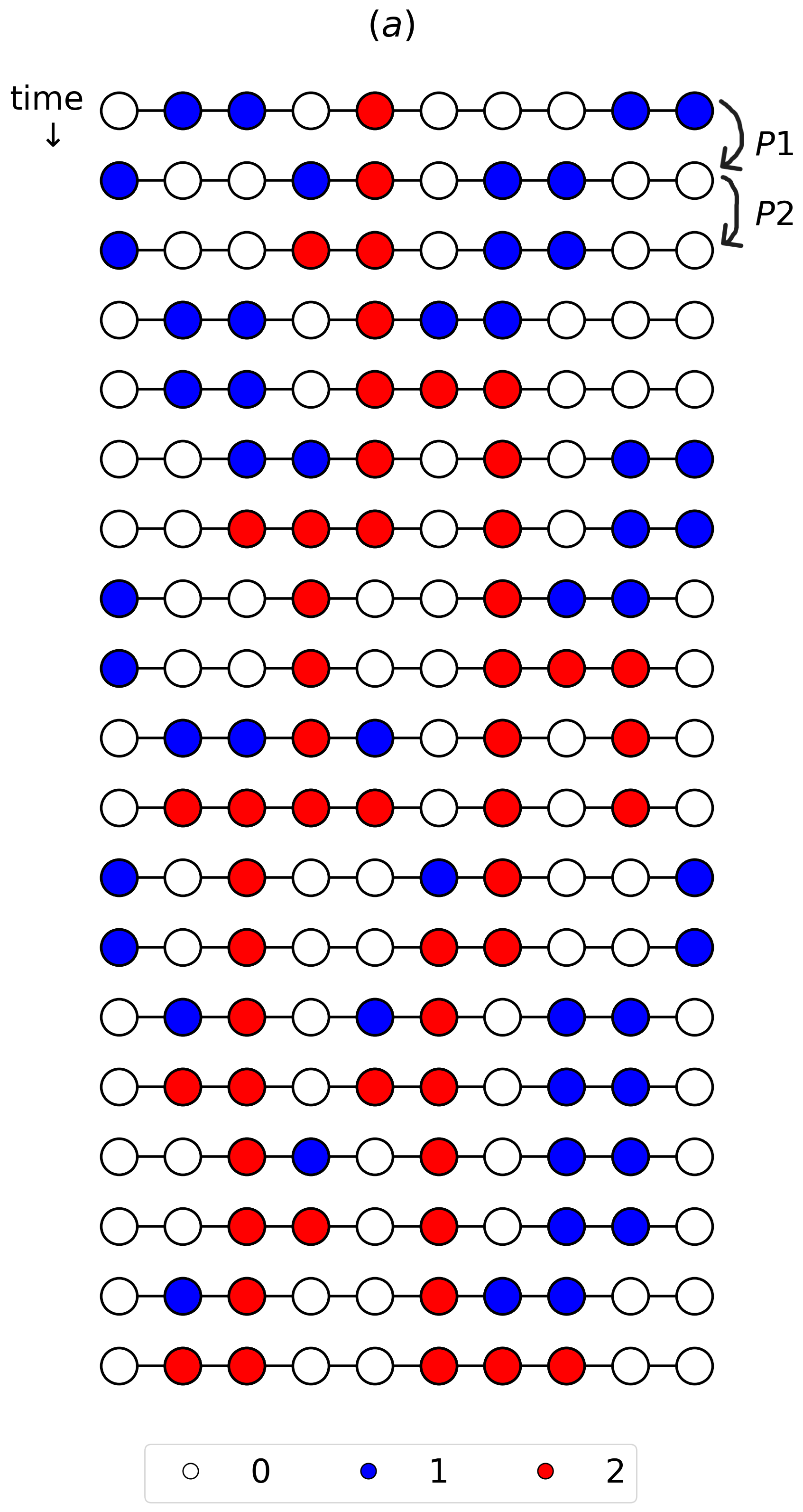}
    \includegraphics[scale=0.78]{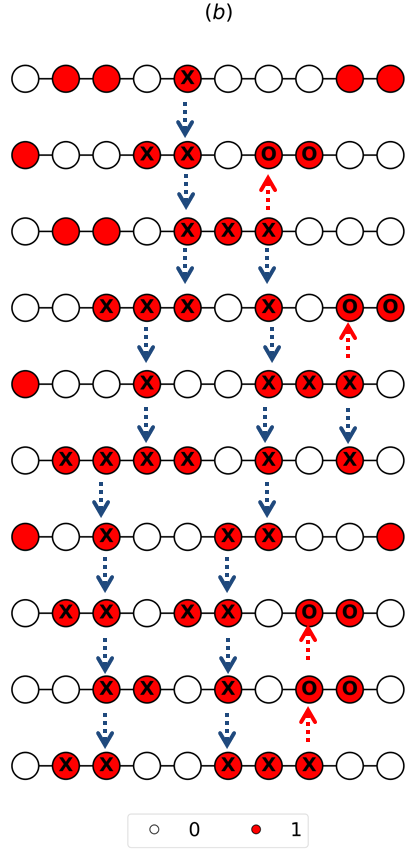}
    \caption{(a) Schematic of the time evolution of sites of a one-dimensional lattice of length $L=10$ in the three species dynamical model of semi-directed percolation. For the sake of clarity, the two processes happening at each time step are represented separately (process P1 defined in Eqs.~\ref{a1},~\ref{a2},~\ref{a3},~\ref{a4}, and P2 in Eq.~\ref{b1}). (b) The active sites alone in a) constitute a semi-directed percolation cluster (sites marked X) on a $10 \times 10$ square lattice with occupation probability $p$. An isotropic cluster includes the additional occupied sites in a) (sites marked O here). Arrows are shown to indicate the \textquoteleft forward\textquoteright\; and \textquoteleft backward\textquoteright\; connections between successive rows on the square lattice.} 
    \label{model_diagram}
\end{figure*}

P2)
\begin{equation}
\begin{aligned}
    1 \rightarrow 2 ~~~~  
    {\textrm{With probability one if it is a part of a}} \\
    {\textrm{contiguous cluster of 1s in contact with a 2.}}
     \label{b1}
\end{aligned}
    \end{equation}
Thus, immune sites become susceptible with probability $p$ and susceptible ones become immune with probability $(1-p)$, whereas they stay in their original states with probability $(1-p)$ and $p$ respectively. Active sites stay active with probability $p$ and become immune with probability $(1-p)$. Finally, activity instantaneously spreads through a cluster of susceptibles in contact with an active site, with immune sites limiting the spread forming boundary of the cluster. A sample progression of the sites of a one-dimensional lattice with the specific dynamical rules defined above for $\epsilon =0$ is shown in Figure.~\ref{model_diagram}.

\begin{figure*}[!ht]
    \centering
    \includegraphics[width=0.5\linewidth]{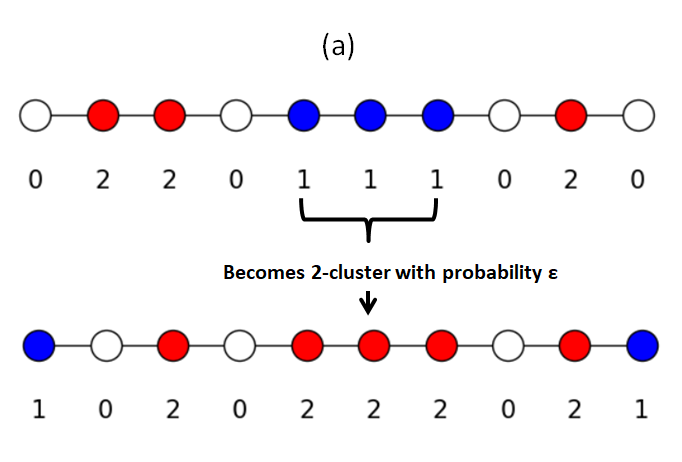}
    \includegraphics[width=0.35\linewidth]{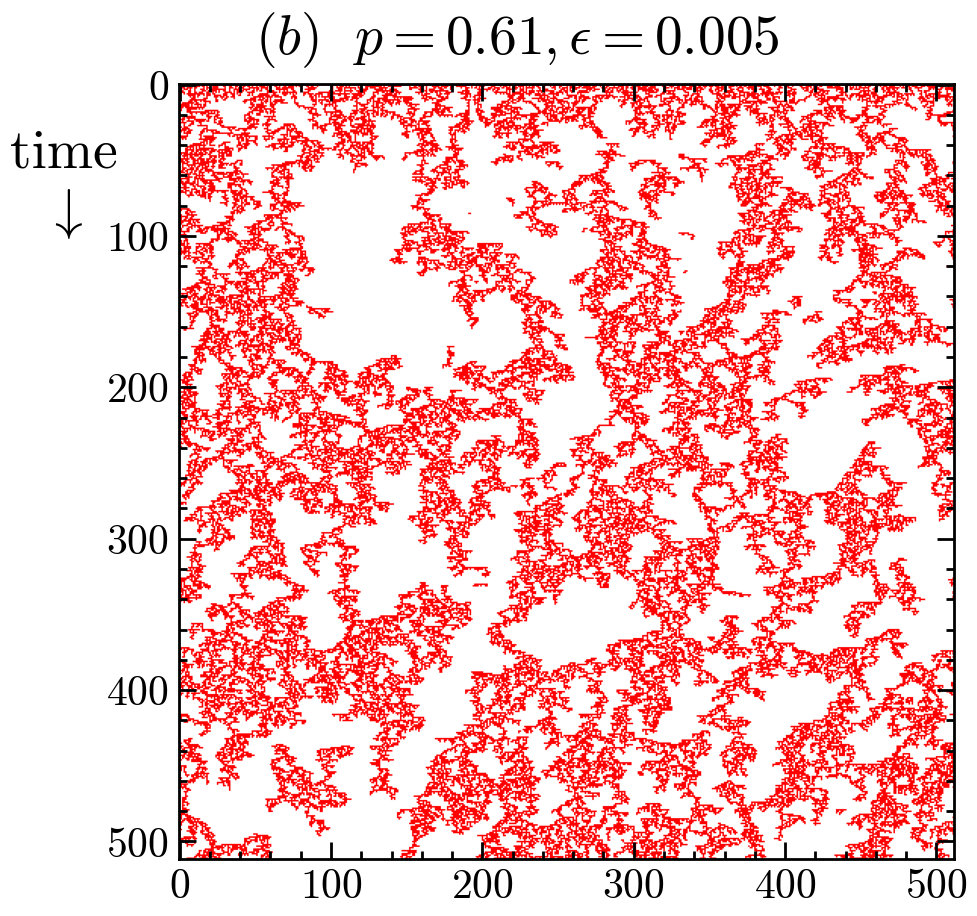}
    \caption{(a) Effect of spontaneous activity is illustrated. Even without the presence of an active neighboring site in state $2$, a $1-$ cluster spontaneously becomes $2-$ cluster with probability $\epsilon$. (b) Typical time evolution of clusters of active sites when spontaneous activity is present. All sites of the one-dimensional lattice are active initially.  Lattice size is $L=512$, $\epsilon=0.005$ and $p=0.61$.}
    \label{config_epsilon}
\end{figure*}

 It is clear from the rules of the dynamics that when $\epsilon = 0$, the state of the system without any activity is absorbing. Starting from an initial configuration with the presence of 2s in an infinite system, activity may die out in a finite number of time steps or sustain forever depending upon the value of the parameter $p$, resulting in an active-absorbing phase transition when $p$ is varied. Note that, in a finite system, for all $p<1$, there is always a nonzero probability that the population of 2s may eventually become extinct. When $\epsilon$ is non-zero, by definition, activity sustains for infinite time in an infinite lattice, thus shifting the phase transition point to the trivial value of $p=0$ (see Figure.~\ref{config_epsilon} for an illustration of dynamics with non-zero $\epsilon$). However, as shown in previous studies of DP with spontaneous activity generation, we may still get behavior reminiscent of a phase transition, termed quasicritical behavior, indicated by the existence of a distinct maximum of susceptibility for non-zero $p$~\cite{Korchinski2021,WilliamsGarcia2014,Fosque2021,Fosque2022,beggs2022,Tian2022}.

From Figure.~\ref{model_diagram}~(a) and (b), we can see that the activity spreads in space (horizontal direction in the figure) and forward in time (downward direction in the figure), but not backwards in time. The horizontal spread of activity in a cluster of 1s is limited by the inactive and immune 0 sites at the boundary of the cluster. Thus, from Figure.~\ref{model_diagram}, we can infer that the spread of activity essentially traces an SDP cluster in time via the dynamic process defined on the one-dimensional lattice. We can also construct such clusters in a non-dynamical way, where after having created a standard square lattice percolation configuration with sites being occupied with probability $p$, a semi-directed cluster is identified (see Figure.~\ref{model_diagram}~(b)). The corresponding isotropic cluster is also shown in the Figure.~\ref{model_diagram}~(b) to illustrate the difference between IP and SDP clusters.

A natural question to ask is, can we define a dynamic 1D model with only two species (instead of three as done here), which will generate SDP clusters in time? It seems that this is not possible. To generate an SDP cluster in time, besides the update rules for each site with time, one must also define a cluster-spreading mechanism that determines the spatial extent of activity spreading in that time slice. This necessitates the introduction of a third species (immune) for marking the boundary of the cluster. In simple terms, to generate an SDP cluster in time, at each time step, check for any surviving active sites; if yes, then determine the extent of its spreading through the susceptible sites, the extent being constrained by the immune sites.

In our problem, With no spontaneous activity present ($\epsilon = 0$), the transition from a regime where the activity persists to the one in which it ceases marks a non-equilibrium phase transition into an absorbing state. In the following, we first study the critical properties associated with the transition for the $\epsilon = 0$ case by Monte-Carlo simulation techniques and thereafter, consider the scenario with spontaneous activity ($\epsilon \neq 0)$. 

Note that the generation of spontaneous activity is contingent on the presence of 1s. Hence, when $p=0$, there will not be any activity present even for non-zero $\epsilon$, which is different from the usual DP models with spontaneous activity~\cite{Korchinski2021}. Also, note that, when $\epsilon=1$, not all sites will become active even for non-zero $p$, making it a non-trivial case.

\section{Results and discussion}
\label{section2}
We simulate the three-species dynamic SDP model defined in Sec.~\ref{section1} on a one-dimensional lattice of size $L$. Periodic boundary condition is employed to reduce finite-size effects. Threshold and exponent estimation are done using a system size of $L=4096$ and $10^3$ independent trials. We will first discuss the pure SDP model ($\epsilon=0$). 

\subsection{Threshold and critical exponents in the dynamic SDP model with \texorpdfstring{$\epsilon =0$}{}}
\label{subsection1}
\begin{figure*}[!ht]
    \centering
    \includegraphics[width=0.33\linewidth]{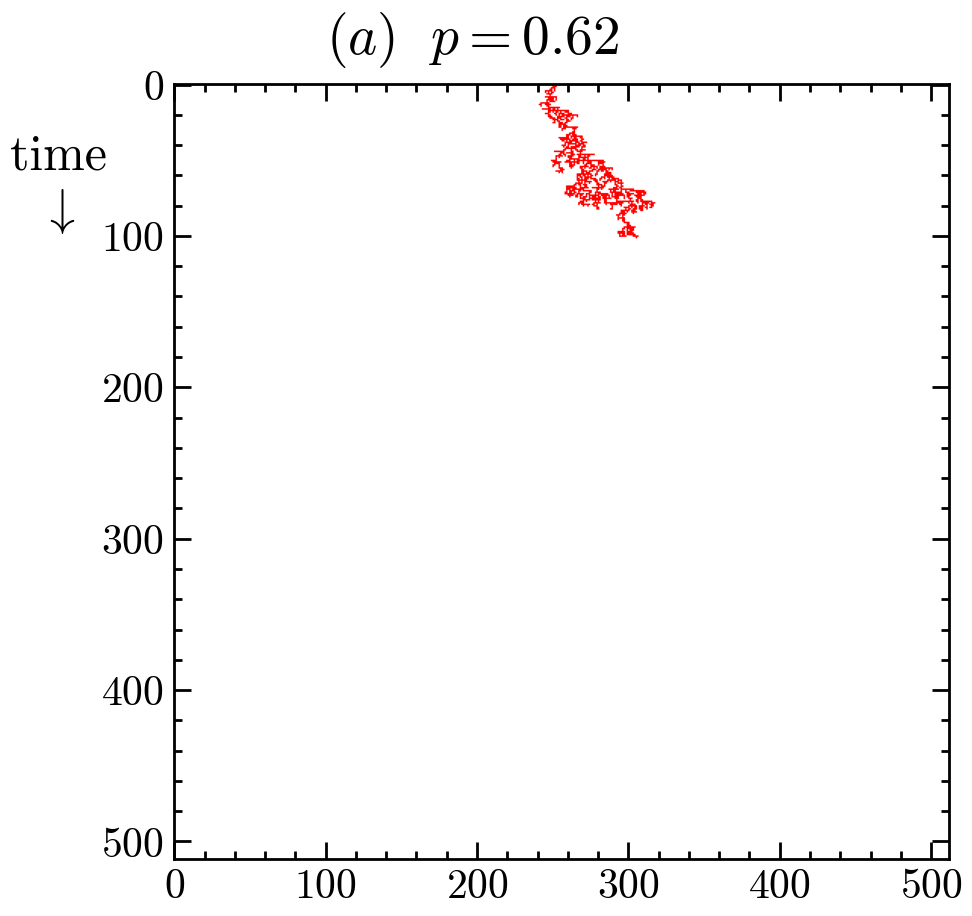}
    \includegraphics[width=0.3\linewidth]{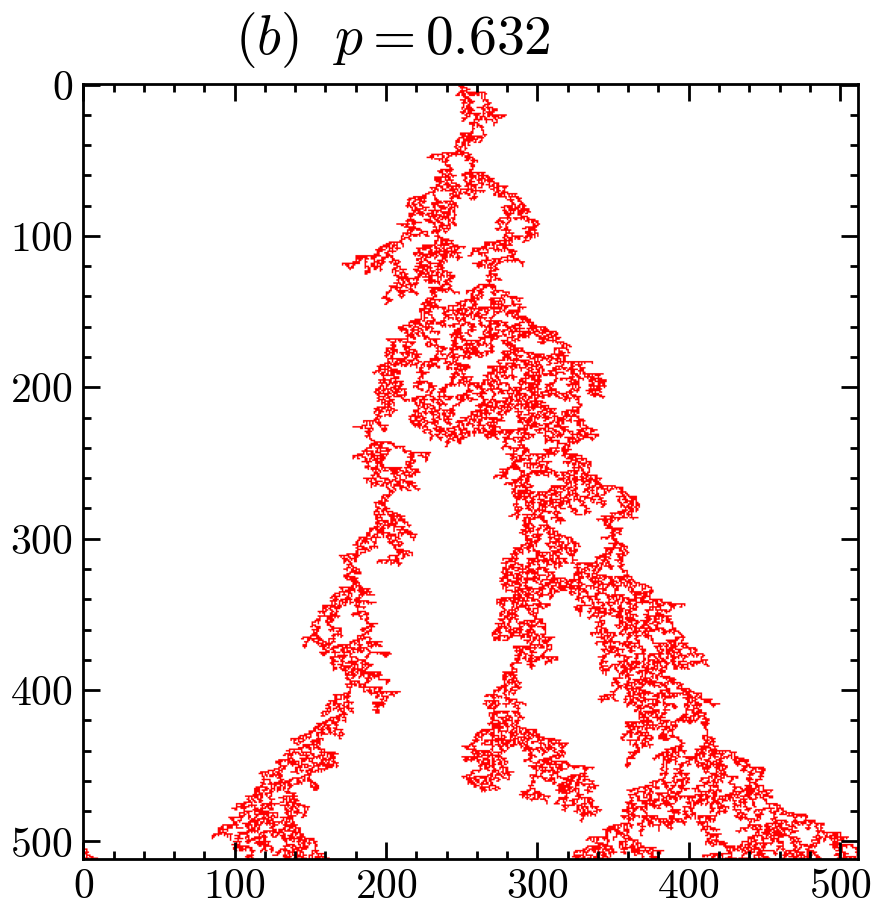}
    \includegraphics[width=0.3\linewidth]{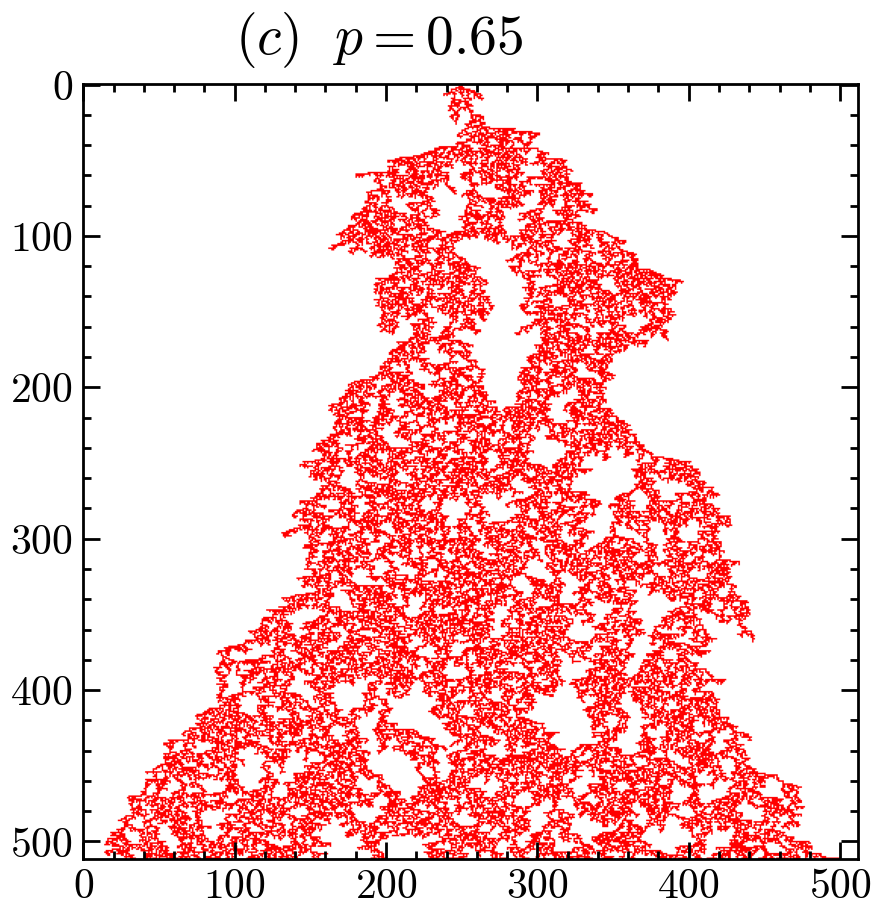}\\
    \includegraphics[width=0.33\linewidth]{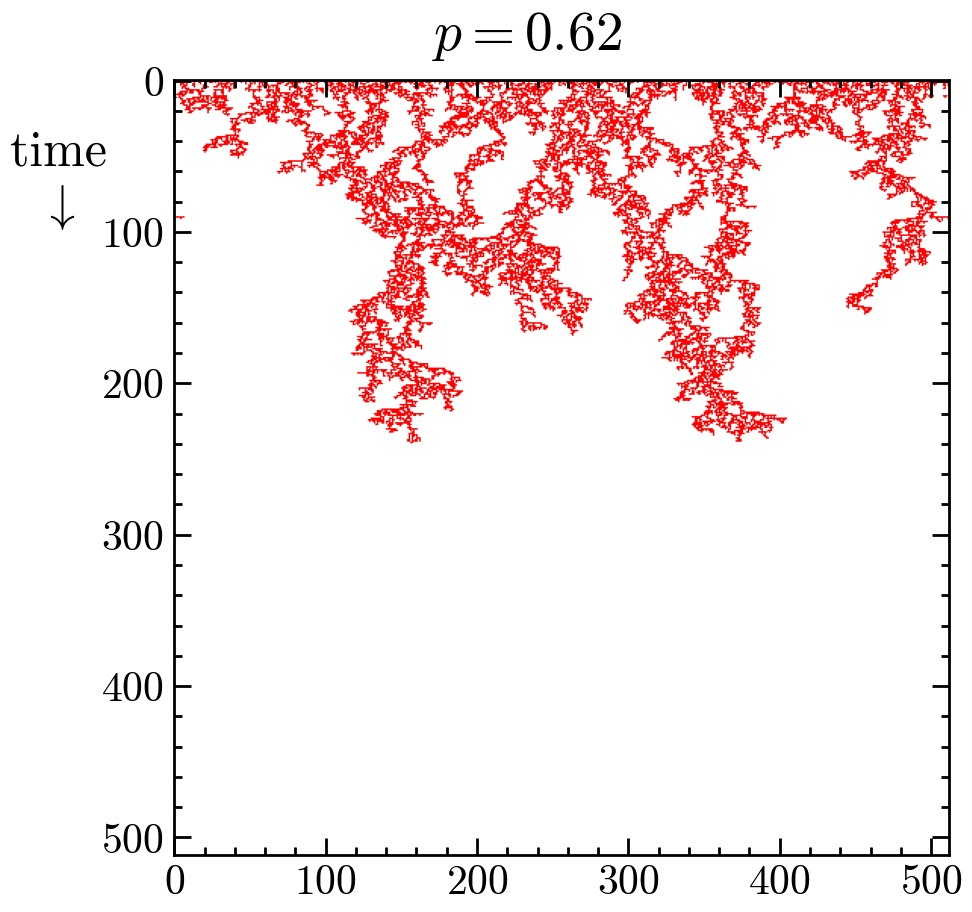}
    \includegraphics[width=0.3\linewidth]{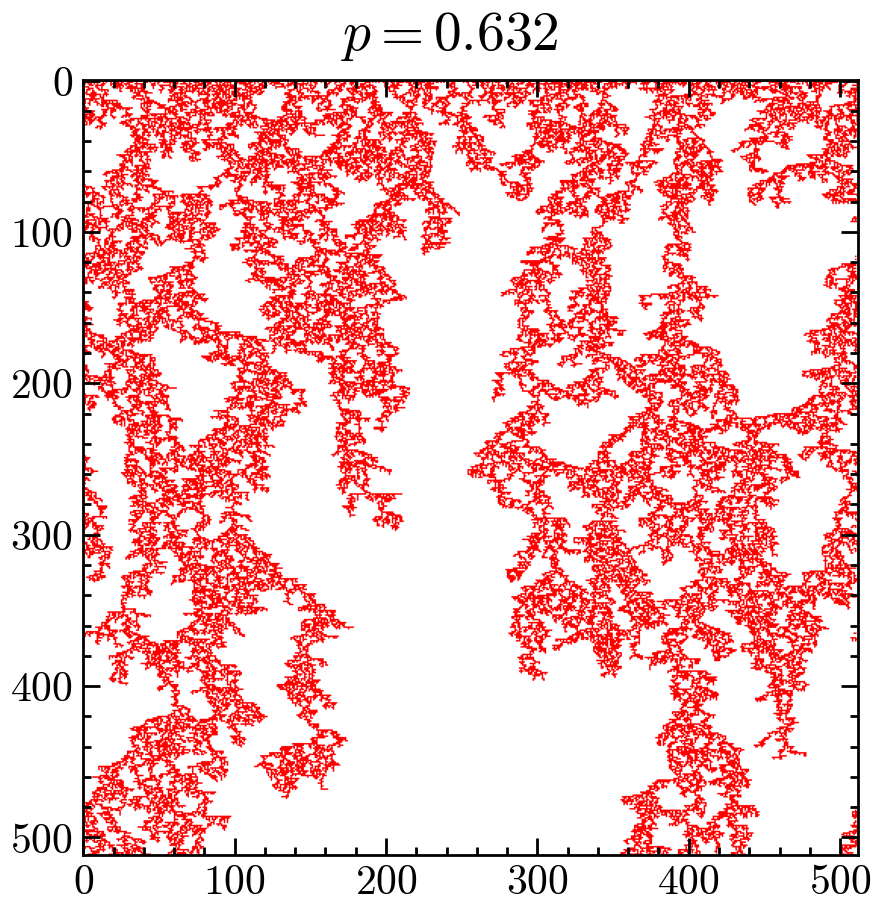}
    \includegraphics[width=0.3\linewidth]{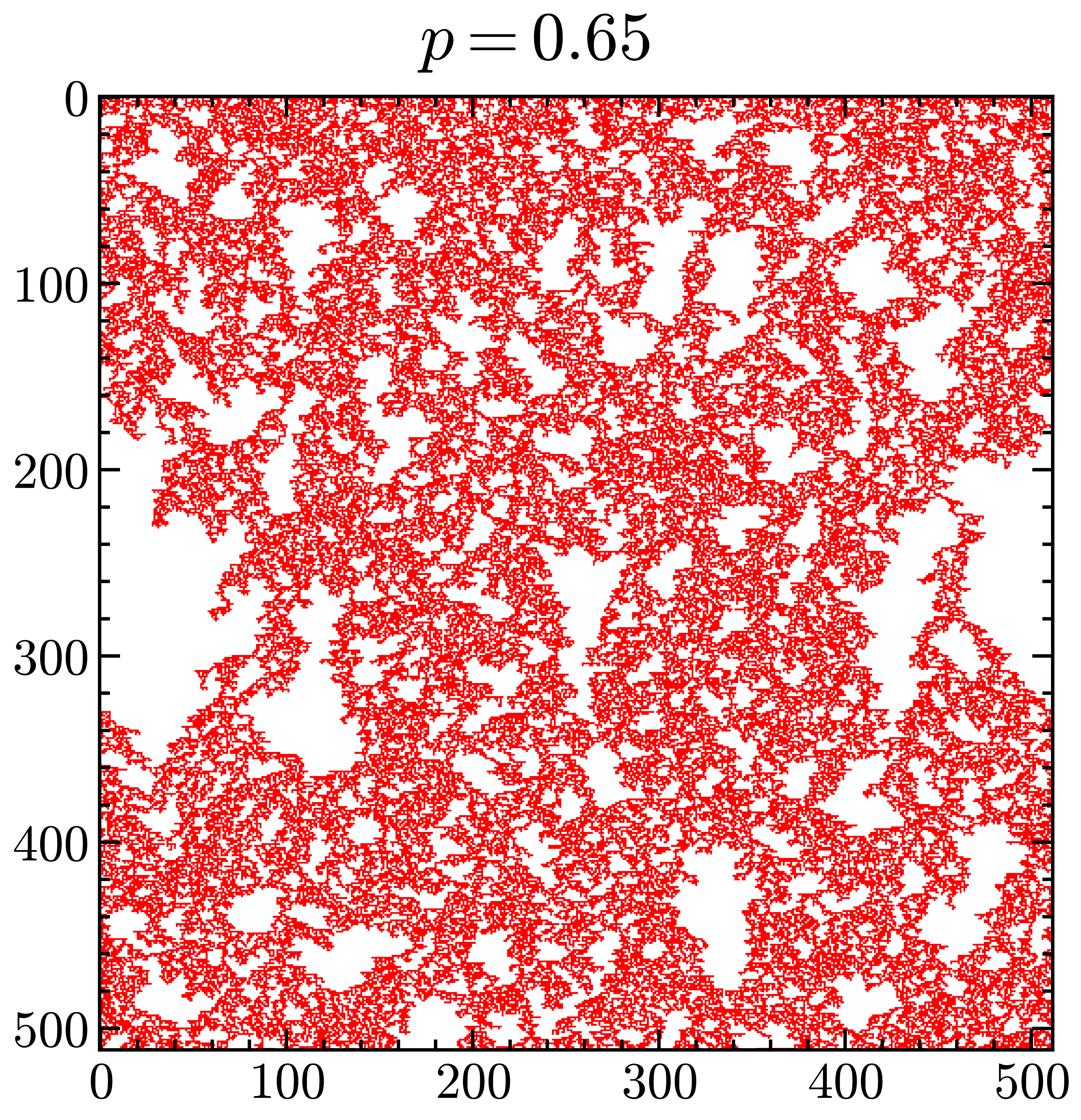}
    \caption{Typical configurations of clusters of active sites generated in the model starting from a single active site (top panel) and $L$ active sites (bottom panel) in a one-dimensional lattice of size $L=512$. Periodic boundary condition is used, and the time evolution of the lattice up to  $512$ time steps is shown. (a) When $p<< p_c$ active cluster dies out very soon. (b) Closer to the threshold, activity survives for longer and longer durations. (c) Well above the threshold $p_c$, activity spreads over the entire system.}
    \label{model_config}
\end{figure*}

We consider two types of initial conditions that are commonly employed when probing the dynamical behavior of systems exhibiting absorbing state phase transitions: a fully active initial configuration and a single active site (or single-seed) at the beginning~\cite{Henkel}. In the present case, the former corresponds to initiating all $L$ sites as $2s$ while the latter corresponds to starting from a single active site at a random location (rest of the sites are $0s$). Representative snapshots of the system dynamics for both the initial conditions are shown in Figure.~\ref{model_config}. We can see that the system's behavior strongly depends on the value of the control parameter $p$. For values of $p$ well below a certain critical threshold, the activity rapidly dies out (Figure.~\ref{model_config}~(a)). Close to the critical value of $p$, the activity exhibits long-term survival (Figure.~\ref{model_config}~(b)), and for $p$ above the threshold, it propagates throughout the entire system (Figure.~\ref{model_config}~(c)).

\begin{figure*}[!ht]
    \centering
    \includegraphics[width=0.49\linewidth]{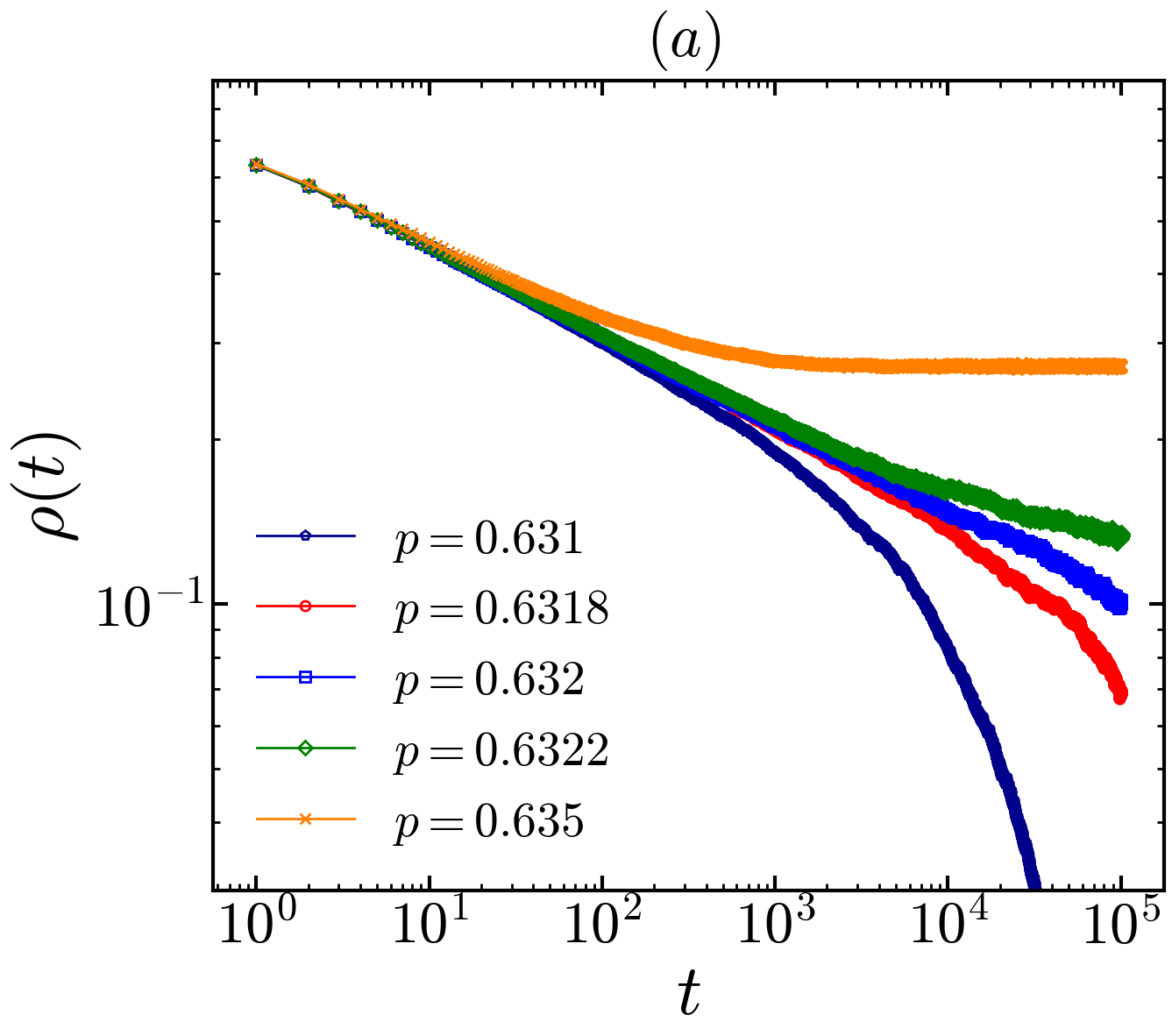}
    \includegraphics[width=0.49\linewidth]{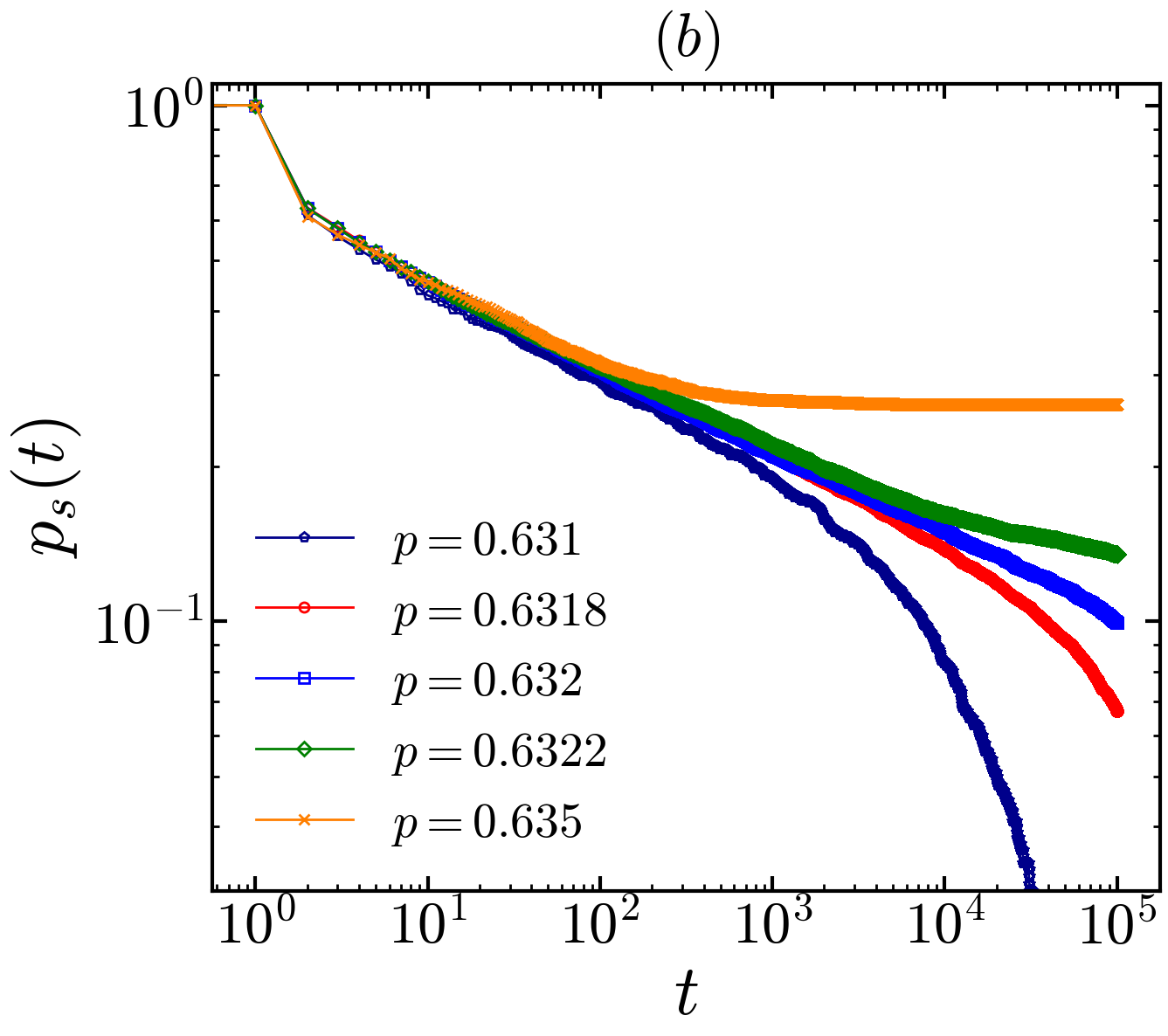}
    \caption{(a) Variation of density of active sites $\rho(t)$ with time $t$ for different values of the parameter $p$ near the threshold. At the threshold $p_c$, we expect a power-law of the form $\rho(t)\sim t^{-\alpha}$. (b) Evolution of survival probability $p_s(t)$ with $t$ for different $p$ values near the threshold. At $p_c$, survival probability obeys a power law of the form $p_s(t) \sim t^{-\delta}$ at threshold.}
    \label{threshold_estimation}
\end{figure*}

For fully active initial configuration, the order parameter is the steady-state value of the average density of active sites, $\rho(t)=\langle \frac{1}{L}\sum_i s_i(t)\rangle$, where $s_i(t) = 1$ if site $i$ is active at time $t$ and $0$ otherwise. Here, angular brackets denote averaging over several trials. For a single seed initial condition, the order parameter is the steady-state value of the survival probability $p_s(t)$, which is the probability of finding at least one active site at time $t$. At the critical point $p_c$, we expect both $\rho(t)$ and $p_s(t)$ to decay as power-laws with time, $\rho(t) \sim t^{-\alpha}$ and $p_s(t)\sim t^{-\delta}$, with critical exponents $\alpha$ and $\delta$~\cite{Henkel}. We can obtain $p_c$ by systematically varying $p$ and determining the interval of $p$ within which we expect a straight line behaviour for $\rho(t)$ and $p_s(t)$ on a log scale. This is shown in Figure.~\ref{threshold_estimation}~(a) and (b), where we can observe that the curves for $p \leq p_1 = 0.6318$ veer downward, indicating an absorbing state, and the curves for $p \geq p_2 = 0.6322$ veer upward, indicating a state above $p_c$, thus yielding $p_c = 0.6320(2)$.
From the slopes of the curves for $p_1$ and $p_2$, we determine the exponent values $\alpha=0.16(1)$ and $\delta=0.16(1)$.

\begin{figure}[!ht]
    \centering
    \includegraphics[width=0.8\linewidth]{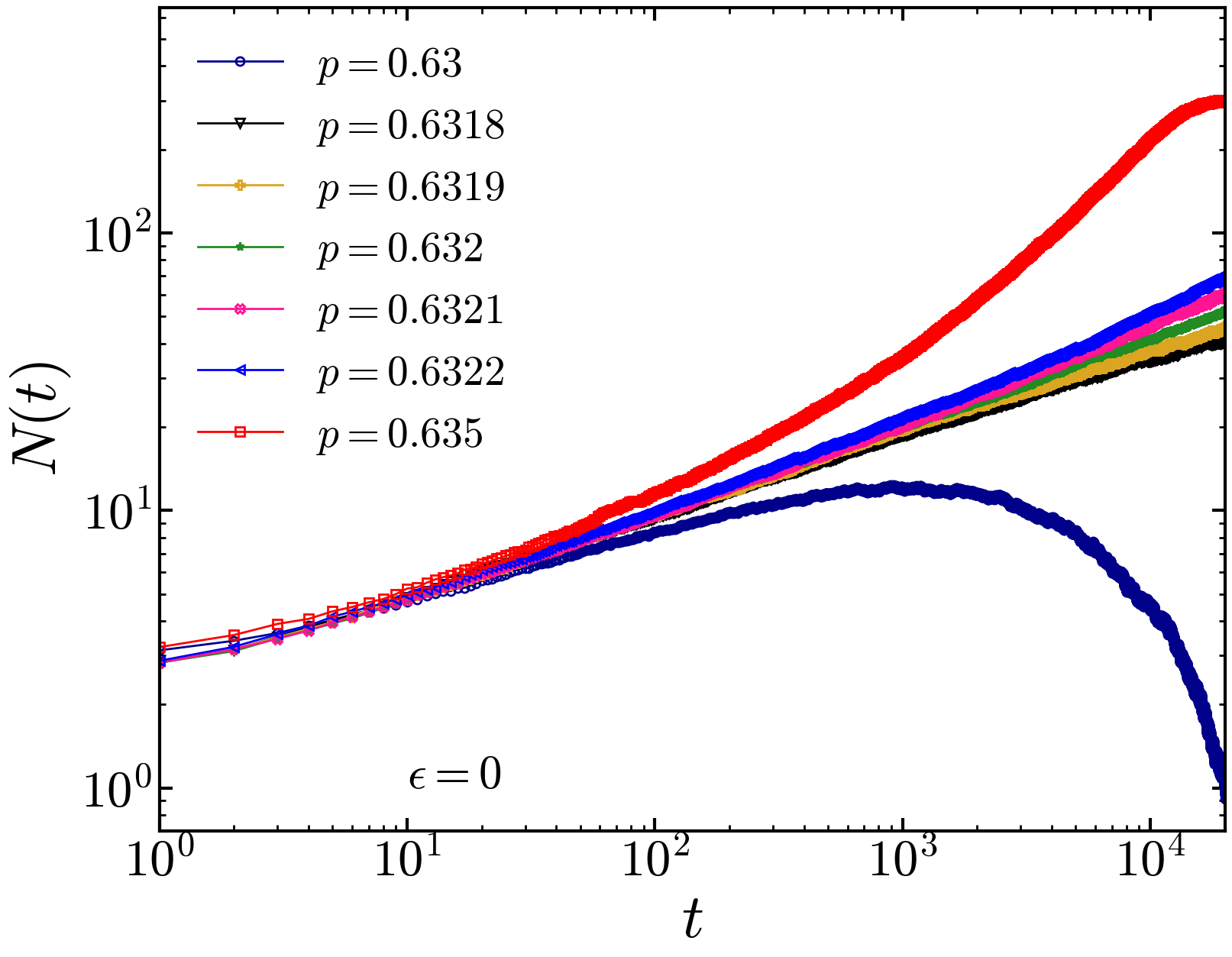}
    \caption{Evolution of the number of active sites $N(t)$ with time $t$ for a system starting with a single active site for different values of the parameter $p$ near the threshold. At $p_c$, $N(t)$ obeys a power law of the form $N(t)\sim t^{\theta}$.}
    \label{Nvst_plot}
\end{figure}

Using the configurations starting from a single seed, an alternate way to obtain the threshold is to determine the number of active sites $N(t)$ at time $t$ as a function of $t$. Again, at the threshold, we expect a power-law behaviour of the form $N(t)\sim t^{\theta}$ with exponent $\theta$.  For $p<p_c$, $N(t)$ decreases exponentially, and for $p>p_c$, $N(t)$ saturates at a constant value. In Figure.~\ref{Nvst_plot}, we show the variation of $N(t)$ with $t$ where we verify the threshold value $p_c = 0.6320(2)$. From the slopes of the straight lines corresponding to $p_1$ and $p_2$, we get the exponent value $\theta=0.31(2)$. The determined threshold value of the dynamical version of SDP is in good agreement with previously obtained thresholds for SDP models. In Ref.~\cite{HOMartin1985}, a $p_c$ value of $0.6317(6)$ was obtained using transfer matrix and phenomenological renormalization techniques and in Ref.~\cite{Knezevic2016}, a value of $0.631985(5)$ was obtained using an improved method.

\begin{figure*}[!ht]
    \centering
    \includegraphics[scale=0.45]{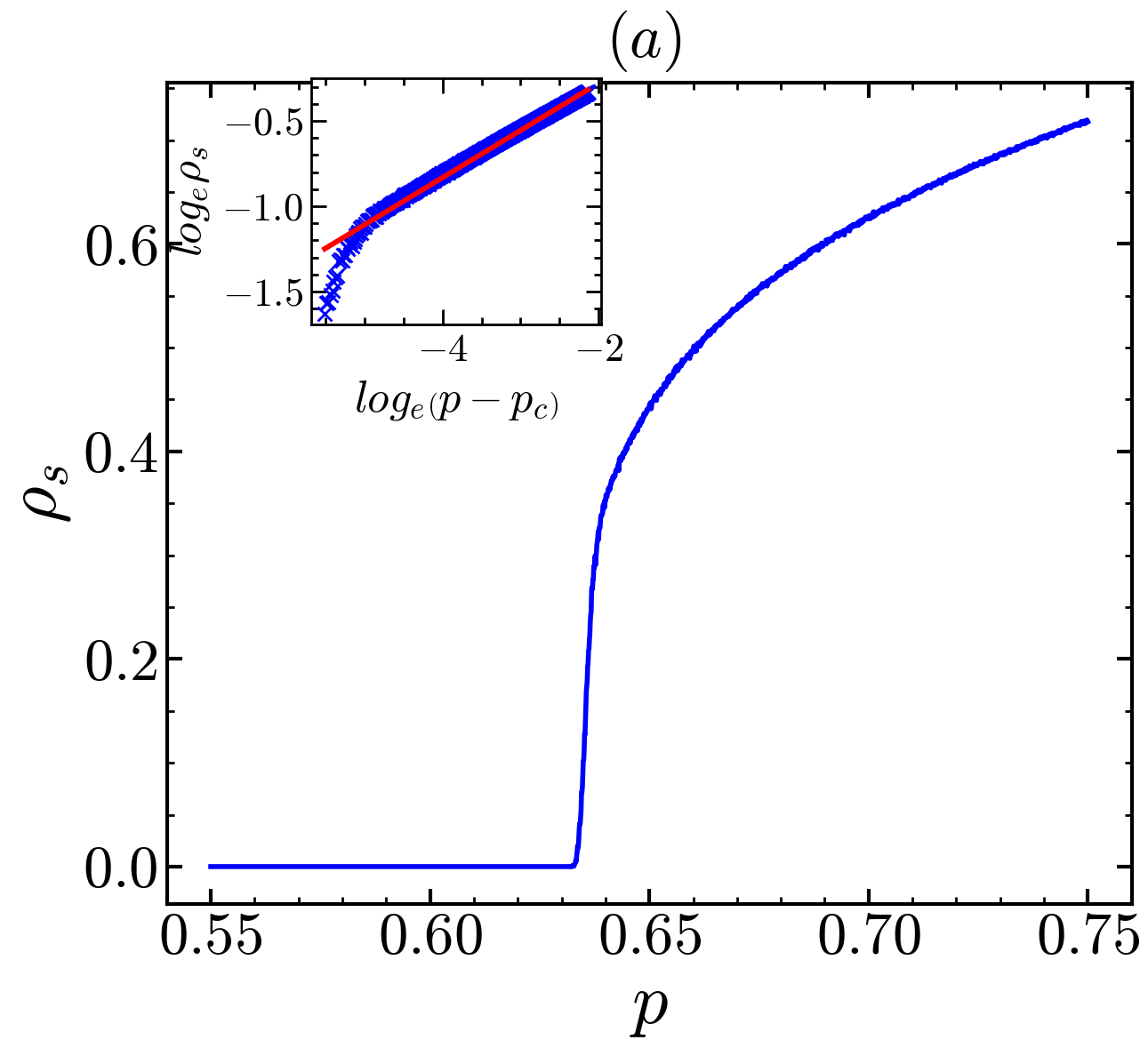}
    \includegraphics[scale=0.45]{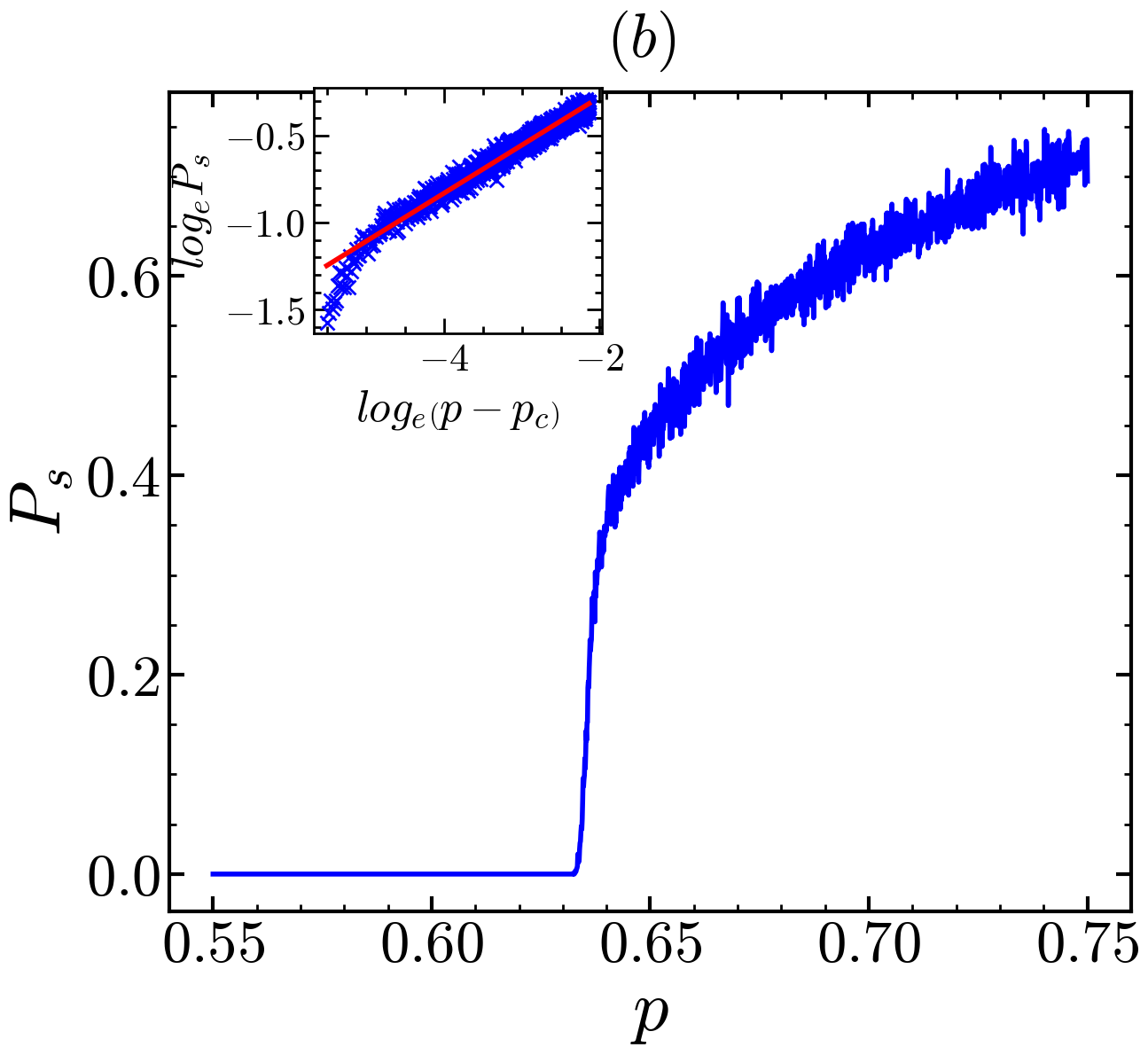}
    \caption{Plots of order parameters against $p$ for fully occupied and single seed initial conditions. (a) Order parameter for the fully occupied case is the density of active sites at the steady state $\rho_s$. Approaching the threshold $p_c$ from above, $\rho_s$ obeys a power law of the form $\rho_s \propto (p-p_c)^{\beta}$ with $\beta$ as the order parameter exponent. The $\log-\log$ plot of $\rho_s$ vs $(p-p_c)$ gives a straight line (shown in inset), the slope of which gives the exponent value $\beta=0.276(1)$. 
    (b) Order parameter for the single seed case is the survival probability at the steady state $P_s$. Approaching the threshold $p_c$ from above, $P_s$ obeys a power law of the form $P_s \propto (p-p_c)^{\beta^{\prime}}$. The $\log-\log$ plot  $P_s$ vs $(p-p_c)$ gives a straight line (shown in inset), the slope of which gives the exponent value $\beta^{\prime}=0.276(1)$. 
    }
    \label{orderparameter}
\end{figure*}

For a fully occupied initial configuration, for $p>p_c$, the density of active sites saturates at a constant value, say $\rho_s$, in the steady-state, and close to the threshold, it follows the power-law form,
\begin{equation}
    \rho_s \propto (p-p_c)^\beta
\end{equation}
where $\beta$ is the order parameter exponent. 
Likewise, for a single seed initial configuration, for $p>p_c$, survival probability saturates at a constant value, say $P_s$, in the steady-state, and close to the threshold follows the power-law form,
\begin{equation}
    P_s\propto (p-p_c)^{\beta^{\prime}}
\end{equation}
with $\beta^{\prime}$ being the associated exponent. Variation of $\rho_s$ and $P_s$ with $p$ is shown in  Figure.~\ref{orderparameter}. The order parameter exponents are obtained from the corresponding $\log-\log$ plots (shown in insets of Figure.~~\ref{orderparameter}). We find that the two exponent values coincide, $\beta=\beta^{\prime}=0.276(1)$, as in the $(1+1)$-dimensional directed bond percolation model, which implies an invariance of cluster structure in the dynamical SDP under time reversal, termed as the rapidity reversal symmetry ~\cite{Henkel}.

\begin{figure*}[!ht]
    \centering
    \includegraphics[width=0.49\linewidth]{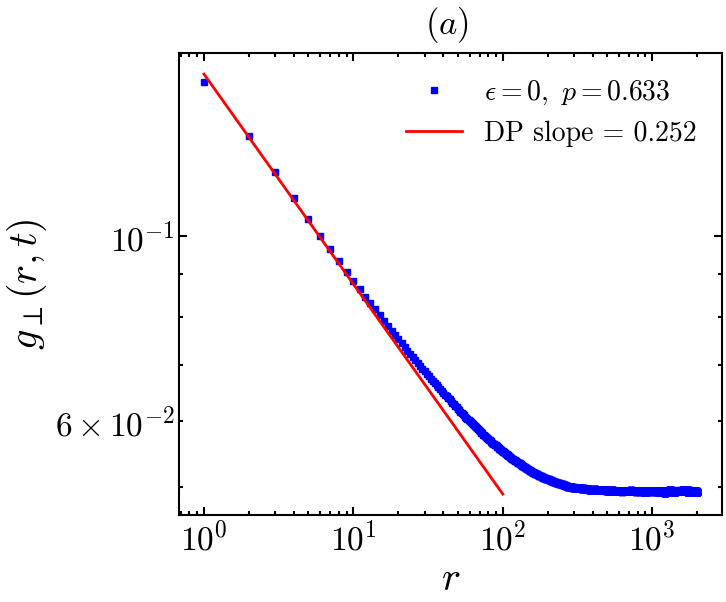}
     \includegraphics[width=0.49\linewidth]{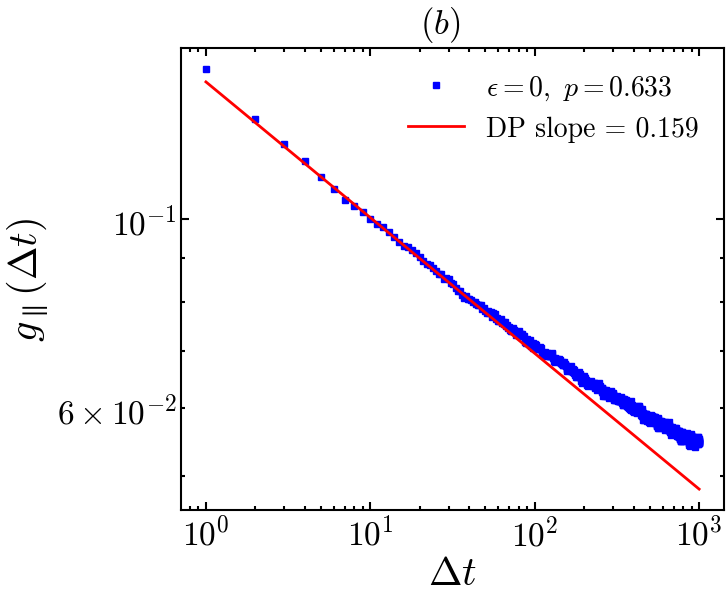}
    \caption{(a) Log-log plot of the equal time correlation function $g_{\perp}(r,t)$ with distance $r$ slightly above threshold $p = 0.633$. DP slope is also shown, which gives the exponent ratio $\beta/\nu_{\perp}$. (b) Log-log plot of the auto correlation function $g_{\parallel}(\Delta t)=\langle s_i(t)s_{i}(t+\Delta t)\rangle$ with time $\Delta t$. DP slope is also shown which gives the exponent ratio $\beta/\nu_{\parallel}$.}
    \label{corr_sdp}
\end{figure*}

Divergence of spatial and temporal correlations at the critical point is a hallmark of critical phenomena. In our problem, the equal-time correlation function $g_{\perp}(r,t)$ is defined as the probability that, at a fixed time $t$, a site located at a distance $r$ from a reference active site is also active and both of them belong to the same active cluster. We can write:
\begin{equation}
    g_{\perp}(r,t)=\langle s_i(t)s_{i+r}(t)\rangle
\end{equation}
where the angular bracket denotes averaging over several pairs of sites $i$ and $i+r$ and also over several trials. Similarly, the autocorrelation function $g_{\parallel}(\Delta t)$ is defined as the probability that an active site at a fixed spatial location is also active after time $\Delta t$ and belong to the same active cluster. We can write:
\begin{equation}
    g_{\parallel}(\Delta t)=\langle s_i(t)s_{i}(t+\Delta t)\rangle
\end{equation}
where the angular bracket denotes averaging over several sites $i$ and also over several trials. Above the critical point, in the steady state, the correlation functions $g_{\perp}(r,t)$ and $g_{\parallel}(\Delta t)$ are expected to decay as $r^{-\beta/\nu_{\perp}}$ and $(\Delta t)^{-\beta/\nu_{\parallel}}$ respectively where $\nu_{\perp}$ and $\nu_{\parallel}$ are the correlation length exponents associated with the divergence of spatial and temporal correlations~\cite{Henkel}. The plots of the spatial and temporal correlation functions are shown in Figure.~\ref{corr_sdp} for a value of $p$ slightly above $p_c$. We verify that the correlations decay as power laws with the DP exponent values $\beta/\nu_{\perp} \approx 0.252$ and $\beta/\nu_{\parallel} \approx 0.159$.

Finally, we consider the dynamic susceptibility $\chi$, which is a measure of the system's response to external stimuli and is related to the variance of activity~\cite{Korchinski2021,WilliamsGarcia2014,Fosque2021,Fosque2022,beggs2022},
\begin{equation}
    \chi=L\left(\langle \rho(t)^2\rangle-\langle \rho(t)\rangle^2\right)
\end{equation}
where $\langle \rho(t)^k\rangle=\frac{1}{T}\sum_{i=1}^{T}\rho(t)^k$ with $T$ as the total time steps considered.

\begin{figure*}[!ht]
    \centering
    \includegraphics[width=0.49\linewidth]{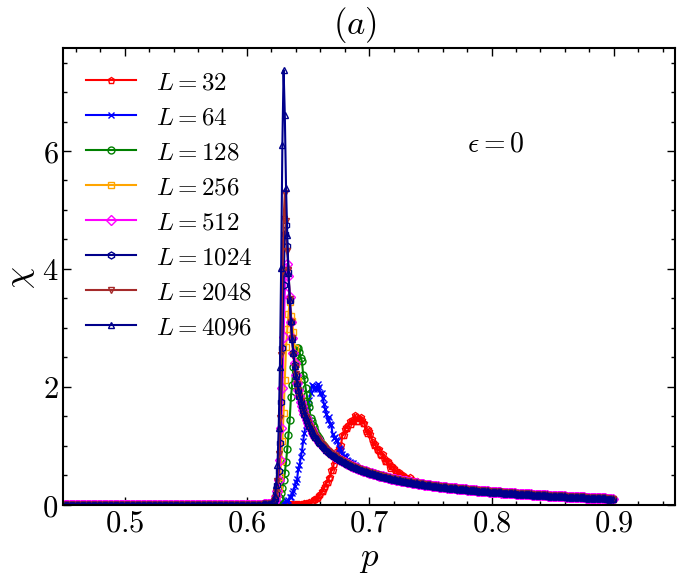}
    \includegraphics[width=0.49\linewidth]{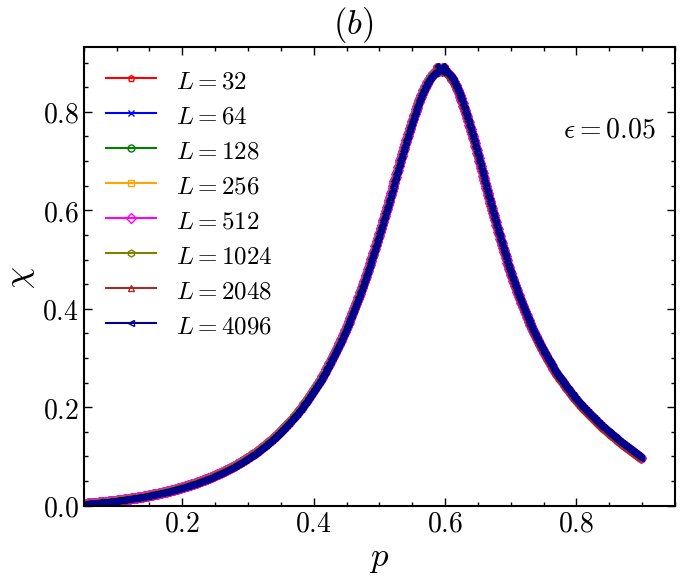}
    \caption{(a) Variation of the dynamical susceptibility $\chi$ with $p$ for different system sizes $L$ for the dynamical SDP with no spontaneous activity ($\epsilon=0$).(b) Variation of the dynamical susceptibility $\chi$ with $p$ for different system sizes $L$ for the dynamical SDP with spontaneous activity (here we use $\epsilon=0.05$).}
    \label{chi_fo}
\end{figure*}

Variation of dynamical susceptibility with $p$ for different values of $L$ is shown in Figure.~\ref{chi_fo}~(a). We can see that the dynamic susceptibility peaks at the threshold with increasing peak heights as we increase $L$, a characteristic behaviour for finite system sizes.

\begin{figure*}[!ht]
    \centering
    \includegraphics[width=0.49\linewidth]{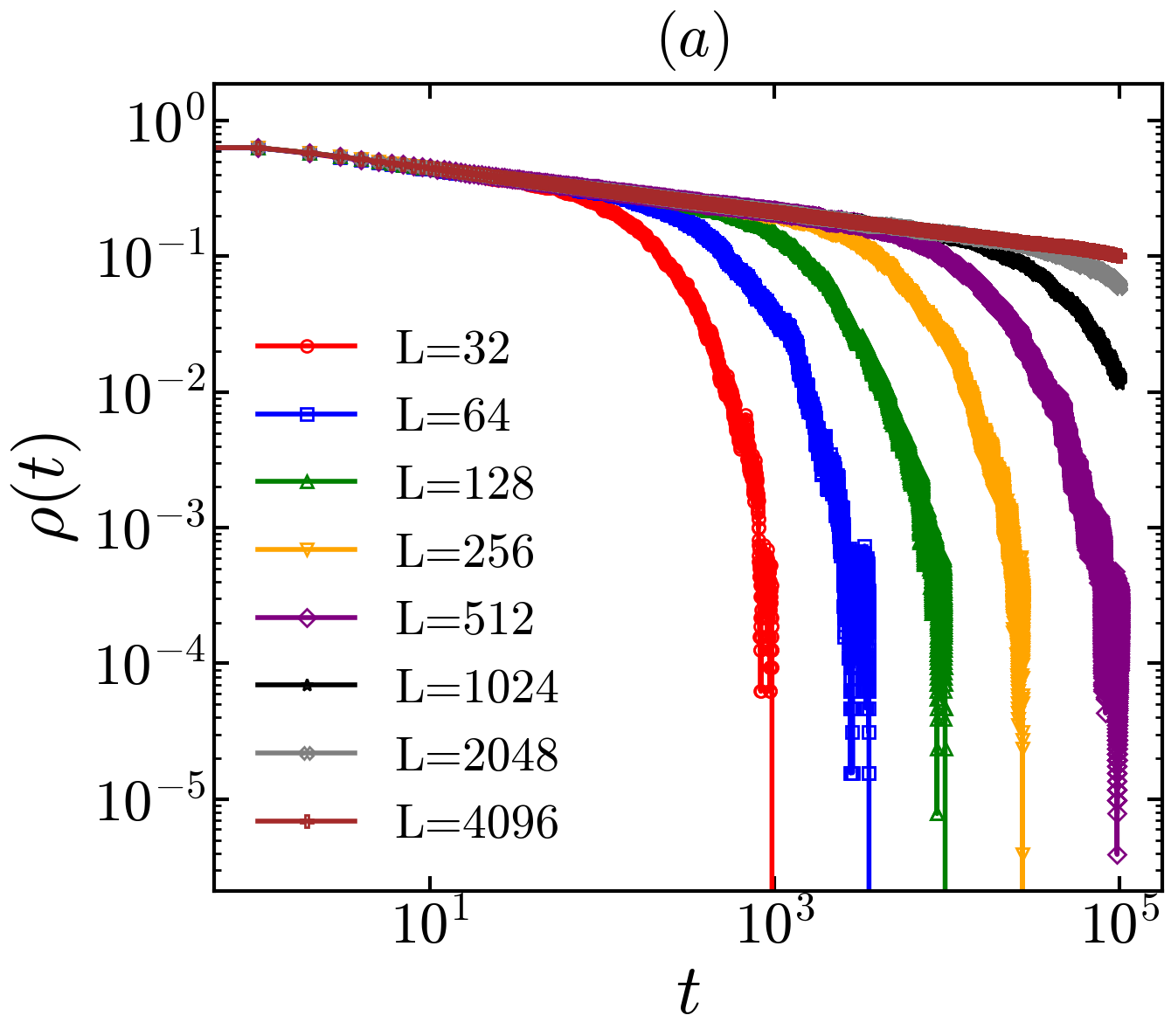}
    \includegraphics[width=0.49\linewidth]{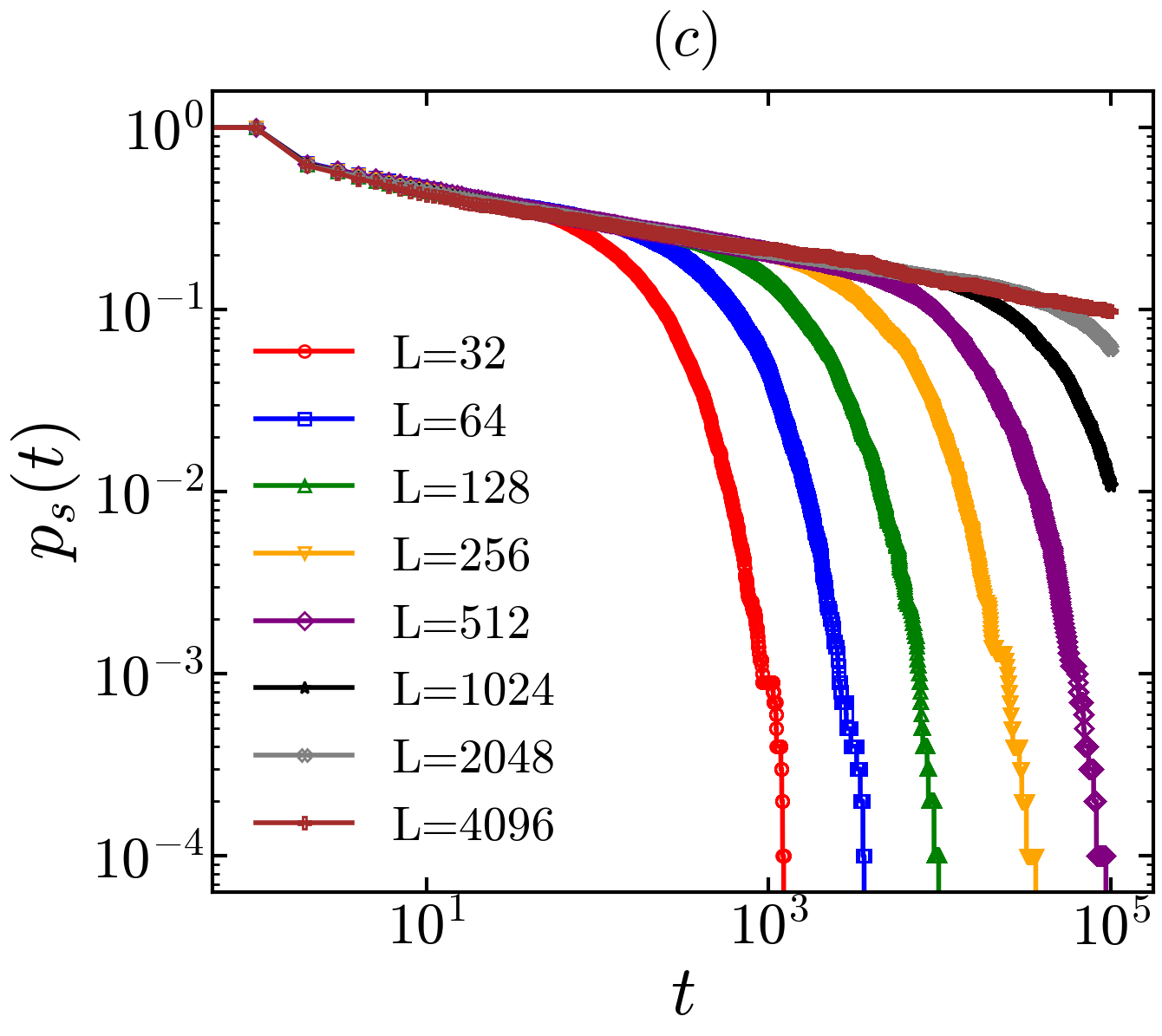}\\
    \includegraphics[width=0.49\linewidth]{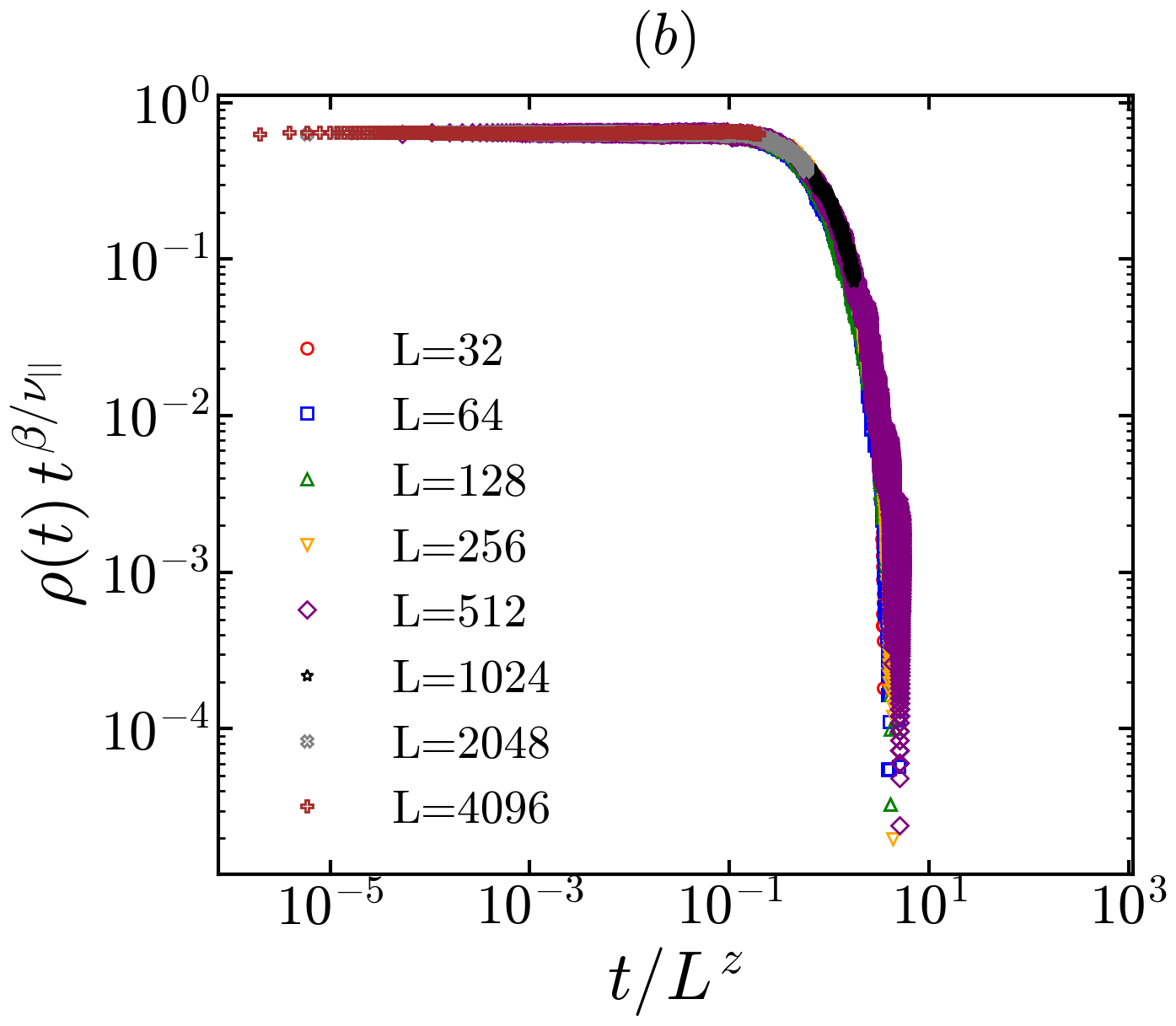}
    \includegraphics[width=0.49\linewidth]{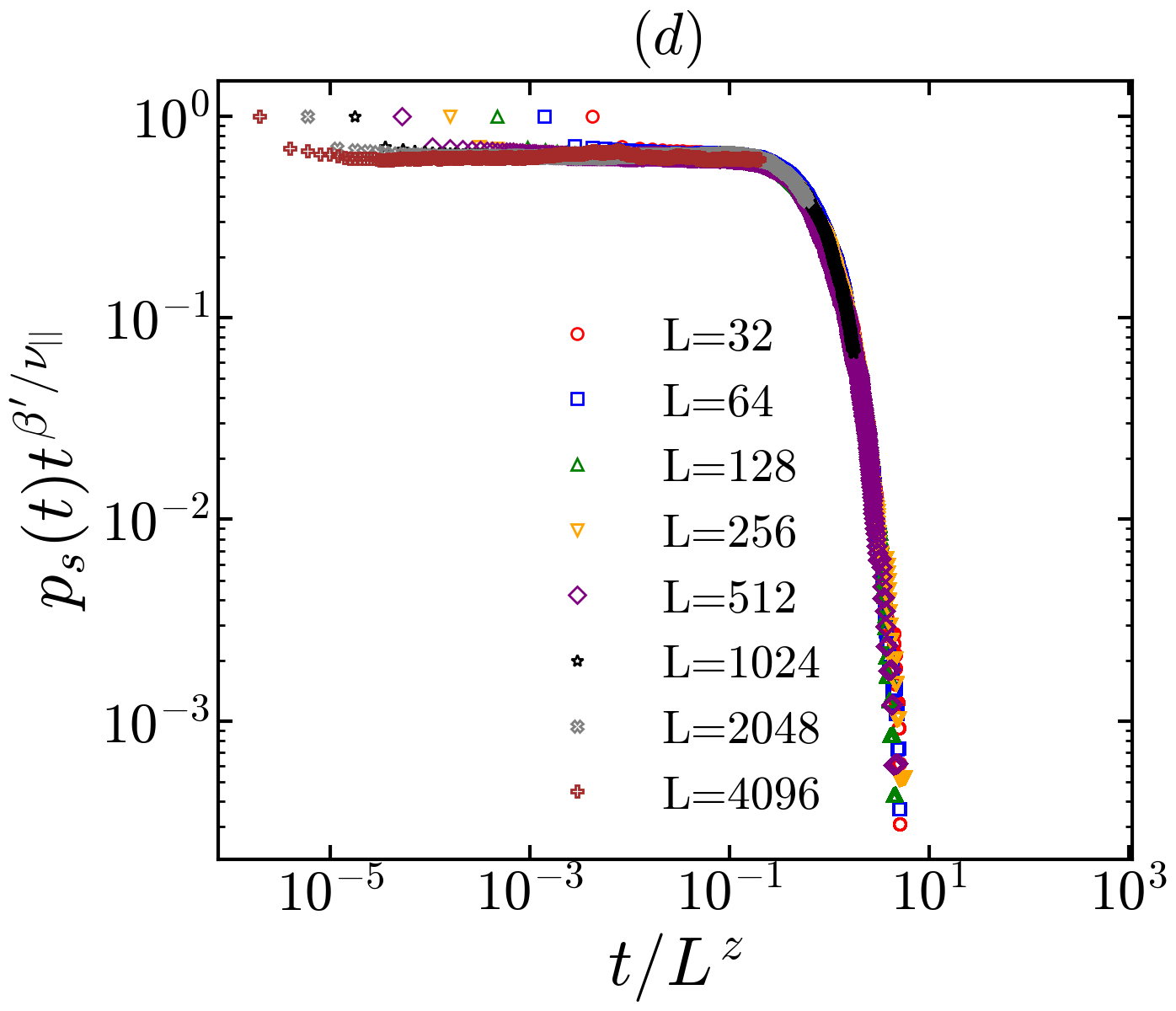}
    \caption{(a \& b) Plot of $\rho(t)$ vs $t$ at the threshold ($p_c\approx 0.632$) for different system sizes and the corresponding data collapse using the DP exponent values $\beta/\nu_{\parallel}= 0.159$ and $z=1.58$ (c \& d) Plot of $p_s(t)$ vs $t$  at the threshold ($p_c\approx 0.632$) for various system sizes and corresponding data collapse with $\beta/\nu_{\parallel}= 0.159$ and $z=1.58$
    }
    \label{datacollapse}
\end{figure*}

\begin{figure*}[!ht]
    \centering
    \includegraphics[width=0.48\linewidth]{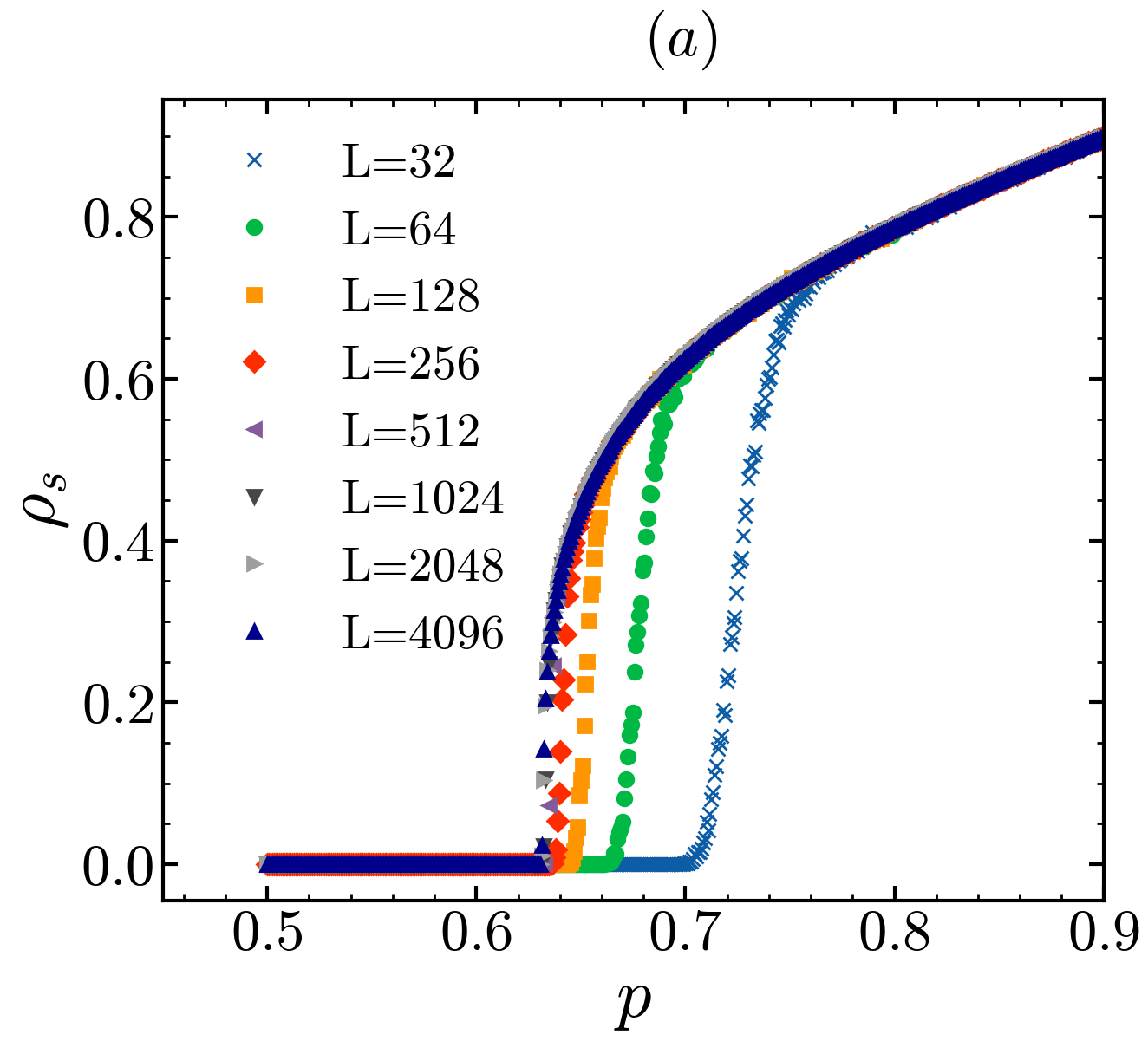}
     \includegraphics[width=0.49\linewidth]{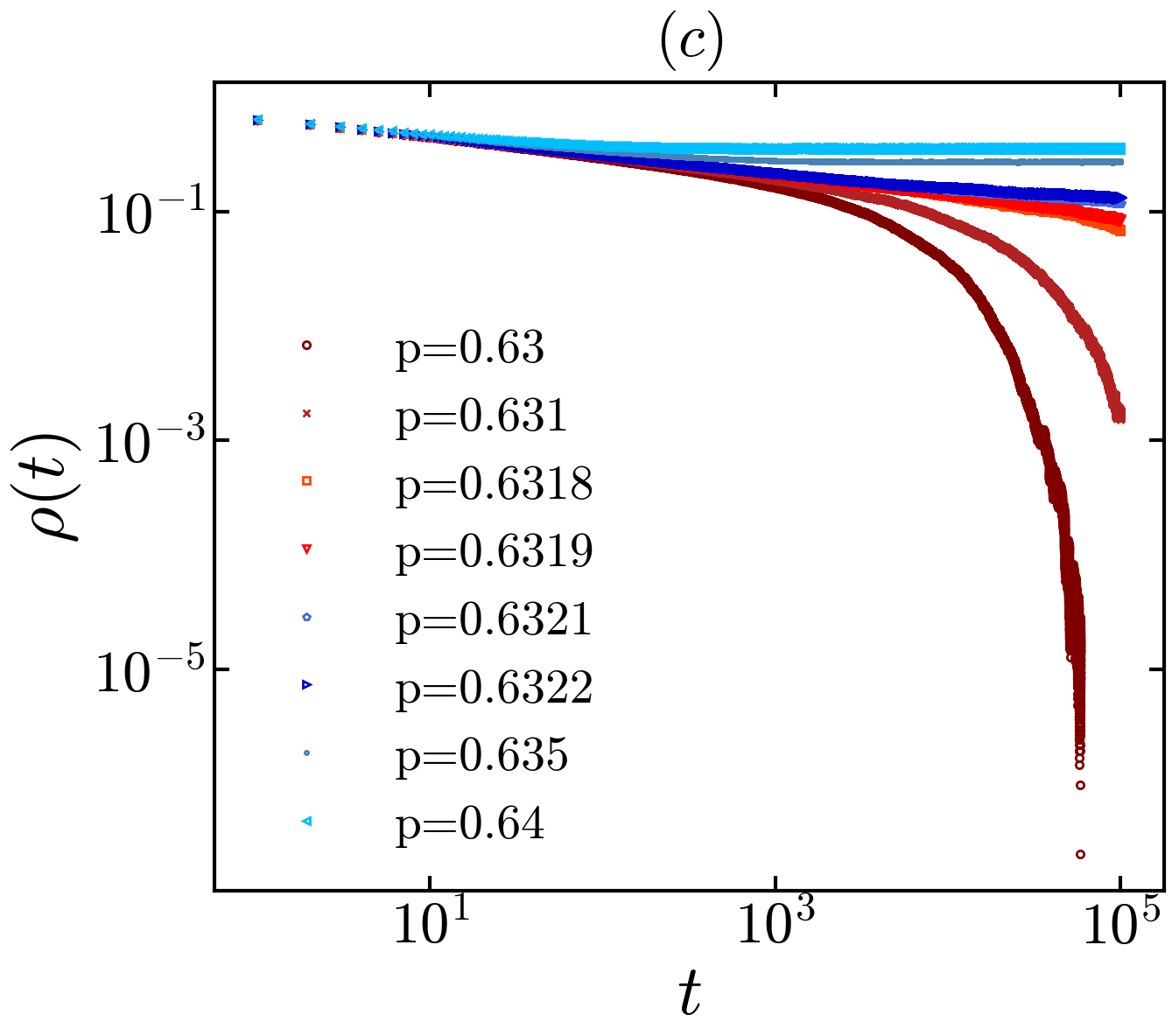}\\
      \includegraphics[width=0.48\linewidth]{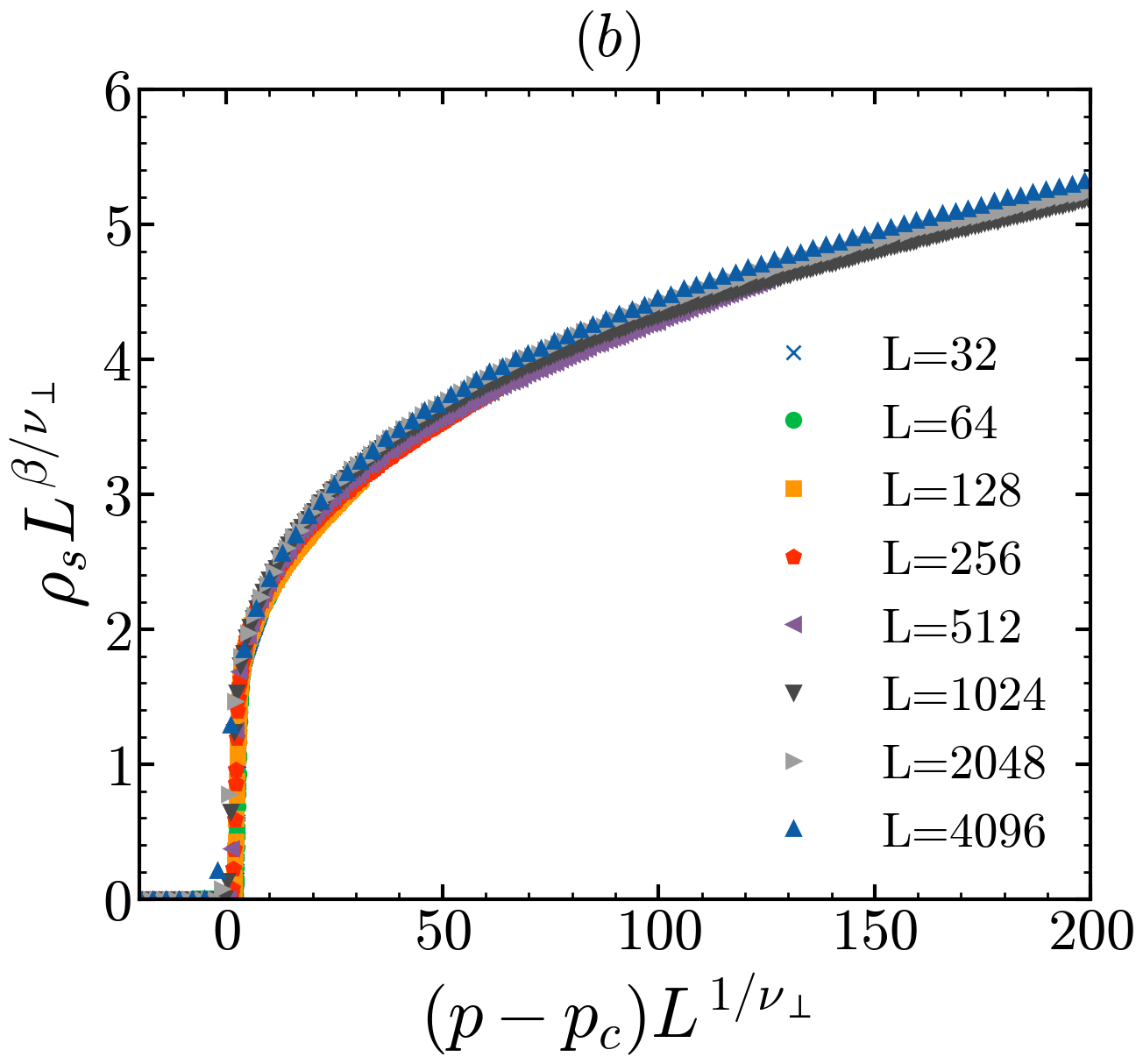}
       \includegraphics[width=0.49\linewidth]{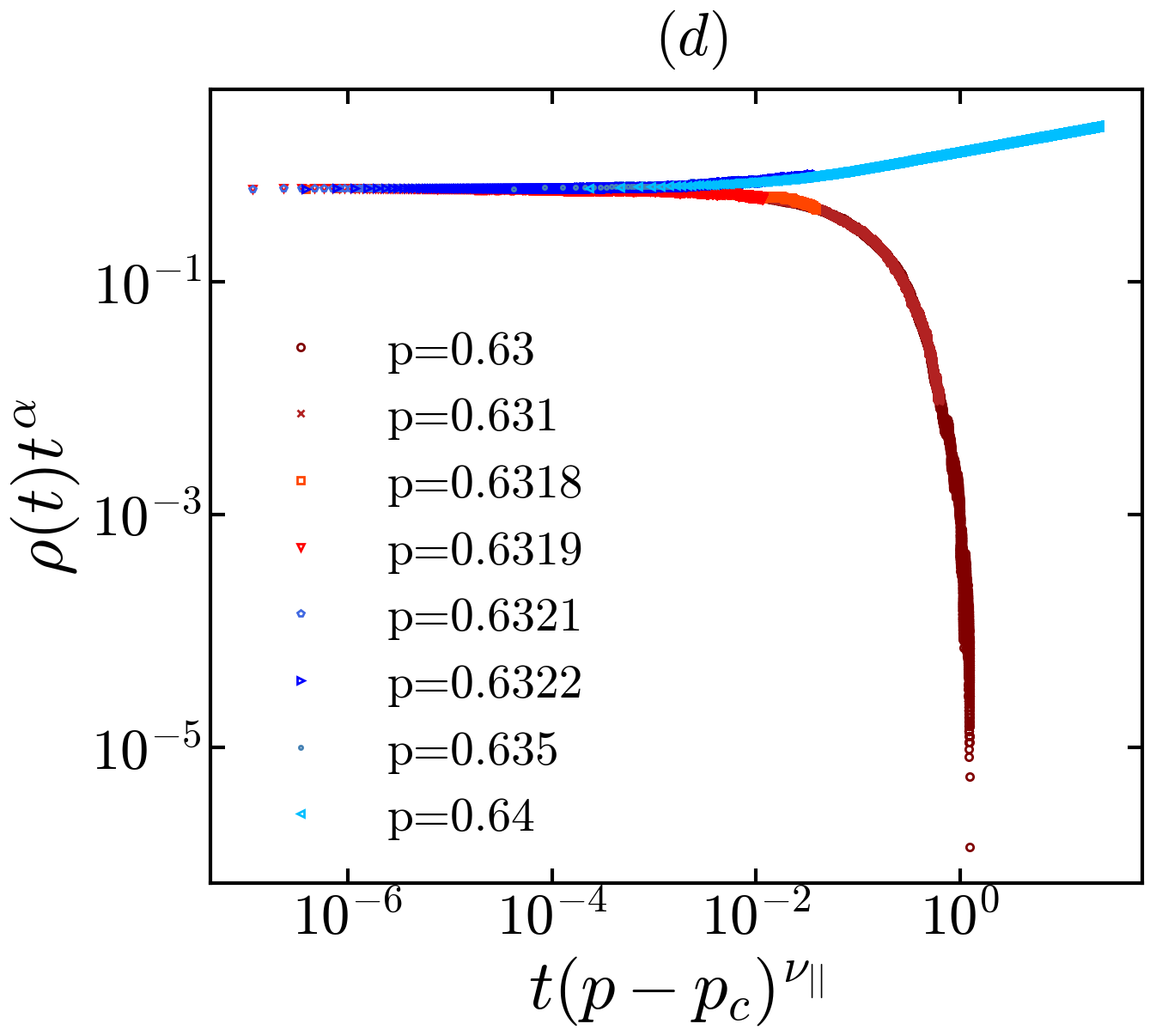}
       \caption{(a \& b) Plot of the steady state density of active sites $\rho_s$ against $p$ for various system sizes and the corresponding data collapse using the DP exponent values $\nu_{\perp}=1.097$, $\beta/\nu_{\perp}=0.159$, and the threshold value $p_c=0.632$. (c \& d) Plot of $\rho(t)$ vs $t$ for different values of $p$ close to and on either side of $p_c$ and the corresponding data collapse. The DP exponent values of $\nu_{\parallel}=1.733$ and $\alpha=0.159$ are used with $p_c=0.632$.}
    \label{nu_datacollapse}
\end{figure*}

The values of the exponents $\beta,\beta',\alpha,\delta, \theta,\beta/\nu_{\perp}$, and $\beta/\nu_{\parallel}$ indicate that the dynamical SDP model belongs to the DP universality class since there is a good agreement between the values \cite{Henkel}. As a further  confirmation of the DP values of the exponents and different finite-size scaling forms, we verify that good data collapses are obtained for the quantities $\rho(t)$, $p_s(t)$, which respectively obey the scaling relations~\cite{Henkel},
\begin{equation}
    \rho(t)=t^{-\beta/\nu_{\parallel}}f_1(t/L^z)
\end{equation}
\begin{equation}
    p_s(t)=t^{-\beta^{\prime}/\nu_{\parallel}}f_2(t/L^z)
\end{equation}
Likewise, near the threshold, $\rho(t)$ and $\rho_s(p,L)$ respectively obey the scaling relations,
\begin{equation}
    \rho_s(p,L)=L^{-\beta/\nu_{\perp}}f_3(L^{1/\nu_{\perp}}(p-p_c))
    \label{scaling3}
\end{equation}
\begin{equation}
    \rho(t)t^{\alpha}= f_4(t(p-p_c)^{\nu_{\parallel}})
    \label{scaling4}
\end{equation}
where $z$ is the dynamical exponent~\cite{Henkel} and $f_i$s are the scaling functions. In Figure.~\ref{datacollapse}~(a) and (b), we plot $\rho(t)$ against $t$ and $\rho(t)t^{\beta/\nu_{\parallel}}$ against $t/L^z$ at $p_c$ respectively for different values of $L$ (32 to 4096) with $p_c = 0.632$ and the DP exponent values $\beta/\nu_{\parallel}=0.159$, $z=1.58$~\cite{Henkel}. In Figure.~\ref{datacollapse}~(c) and (d), we plot $p_s(t)$ against $t$ and $p_s(t)t^{\beta/\nu_{\parallel}}$ against $t/L^z$ at $p_c$ respectively for different values of $L$. We can see that good data collapses are obtained. Likewise, the scaling forms given by Eqs.~\ref{scaling3} and \ref{scaling4} lead to good data collapse curves as depicted in Figure.~\ref{nu_datacollapse} when we use the DP exponent values $\beta= 0.276, \nu_{\perp}=  1.097, \nu_{\parallel} = 1.733$, and $\alpha = 0.159$.

\subsection{Dynamical SDP with spontaneous activity}
\label{subsection2}
When $\epsilon > 0$, by definition, activity will not go extinct for any non-zero value of $p$ in an infinite system, and hence there will not be any phase transition. Although there is no distinct critical point in such a case, there could still be non-trivial behaviour exhibited by quantities like susceptibility and correlations, such as a maximum in the former and power-law decays in the latter at some value of the control parameter $p$ termed as the pseudo threshold~\cite{Korchinski2021}.

\begin{figure}[!ht]
    \centering
    \includegraphics[width=0.8\linewidth]{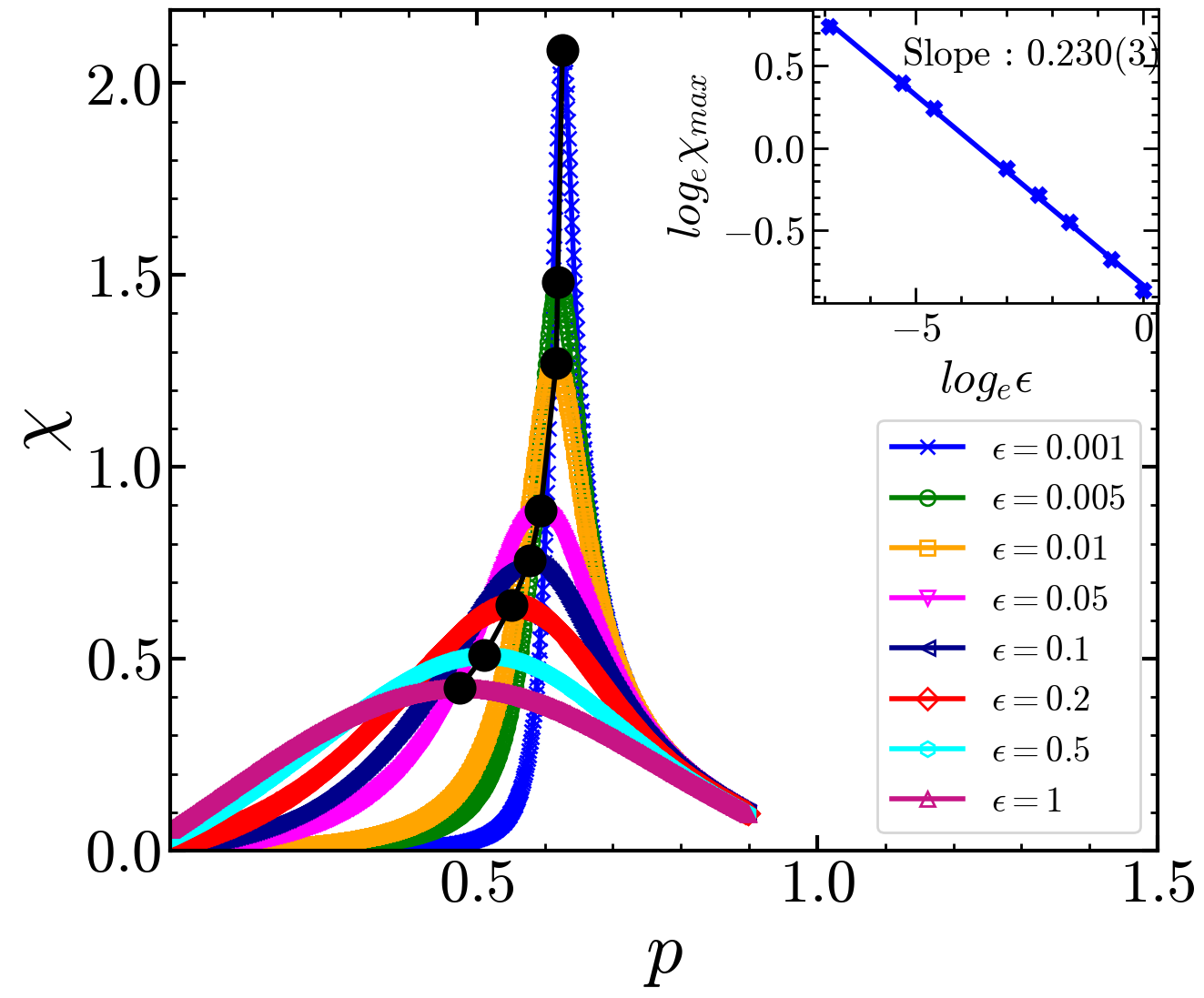}
    \caption{Variation of dynamic susceptibility $\chi$ with $p$ for different values of $\epsilon$. For each $\epsilon$, peak of susceptibility is marked in black circles. The line joining the peaks is said to be a widom line. (inset) Log-log plot of the peak susceptibility $\chi_{max}$ vs $\epsilon$ with the best straight line fit.}
    \label{widom}
\end{figure}

In Figure.~\ref{chi_fo}~(b), we show the variation of the dynamic susceptibility with $p$ for different $L$ for $\epsilon = 0.05$. We can see that susceptibility plots show a maximum for a particular value of $p$ independent of $L$. We can contrast this with the behaviour for $\epsilon = 0$ seen in Figure.~\ref{chi_fo}~(a), where the susceptibility indicates a diverging behavior for larger values of $L$ as expected for the usual critical behaviour. This indicates the existence of a \emph{pseudo-threshold} and associated quasi-critical behaviour for non-zero values of $\epsilon$. In Figure.~\ref{widom}~(a), we show the dynamic susceptibility plots for different values of non-zero $\epsilon$. We can see that the value of $p$ at which the peak of $\chi$ is obtained shifts to lower values as we increase $\epsilon$. Also, the peak height decreases and curves become broader as we increase $\epsilon$. The broadening of the susceptibility peak indicates that, rather than a specific critical point, the system transitions through a broader critical-like region depicting a change from sharp criticality at $\epsilon = 0$ to quasi-critical dynamics for $\epsilon >0$~\cite{Korchinski2021,WilliamsGarcia2014,Fosque2021,Fosque2022,beggs2022}.  The curve obtained by joining the points of maximal susceptibility in Figure.~\ref{widom} defines a non-equilibrium Widom line, similar to the findings in systems with underlying spontaneous activation \cite{Korchinski2021,Fosque2021,Fosque2022}.
The variation of the peak values of $\chi$ with $\epsilon$ follows a power law of the form $\chi_{max} \sim \epsilon^{a}$ with $a = 0.230(3)$ as shown Figure.~\ref{widom} (inset). 

\begin{figure*}[!ht]
    \centering
    \includegraphics[width=0.49\linewidth]{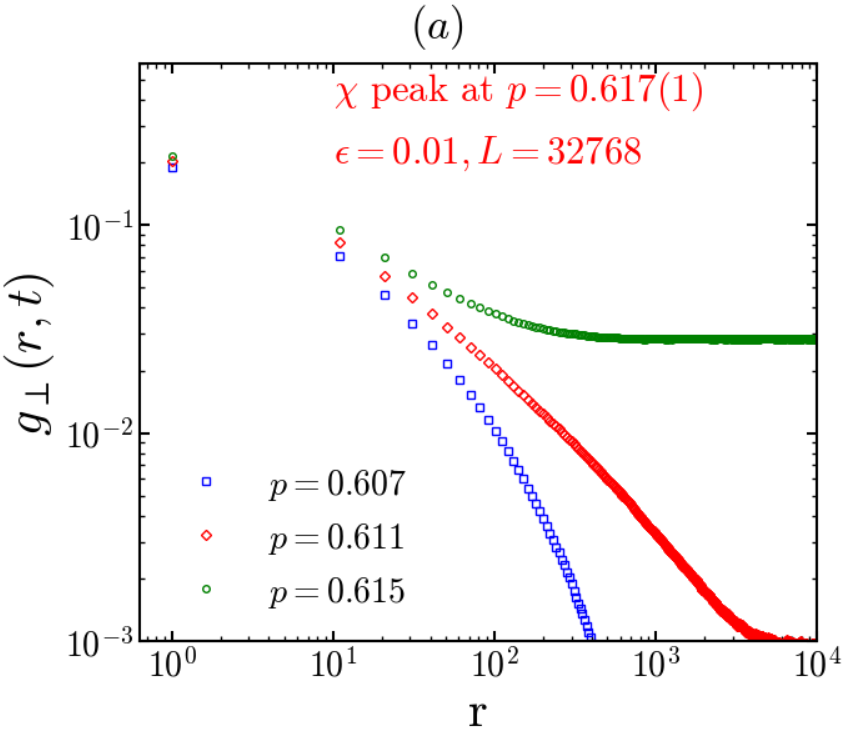}
    \includegraphics[width=0.49\linewidth]{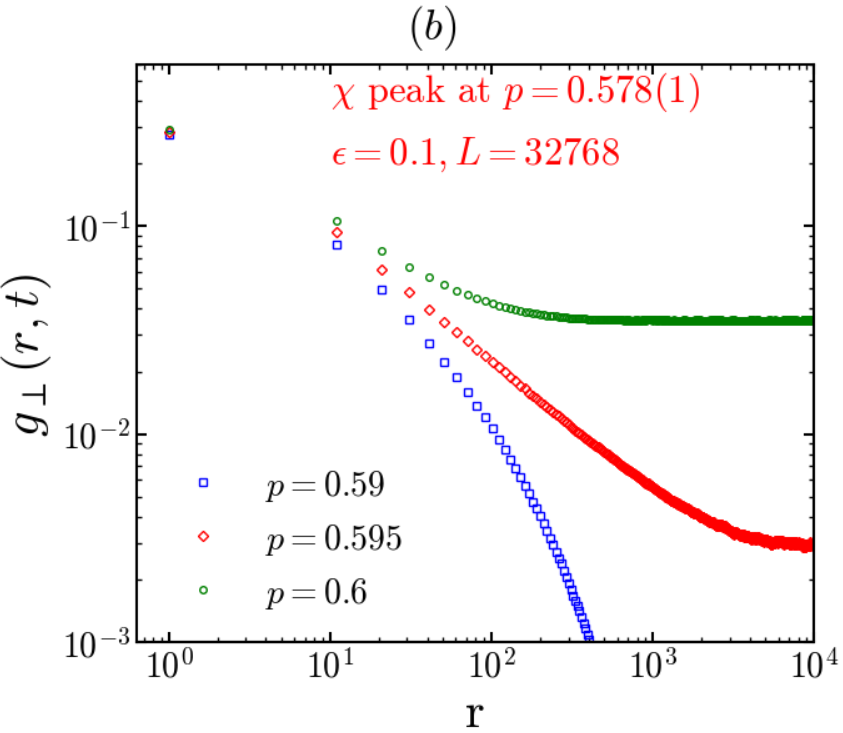}\\
    \includegraphics[width=0.49\linewidth]{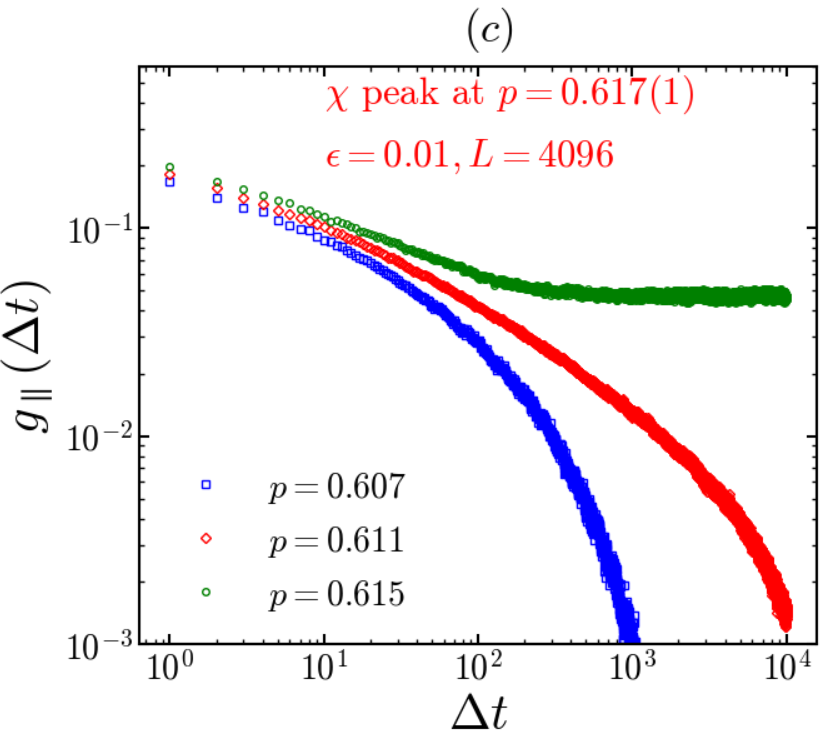}
    \includegraphics[width=0.49\linewidth]{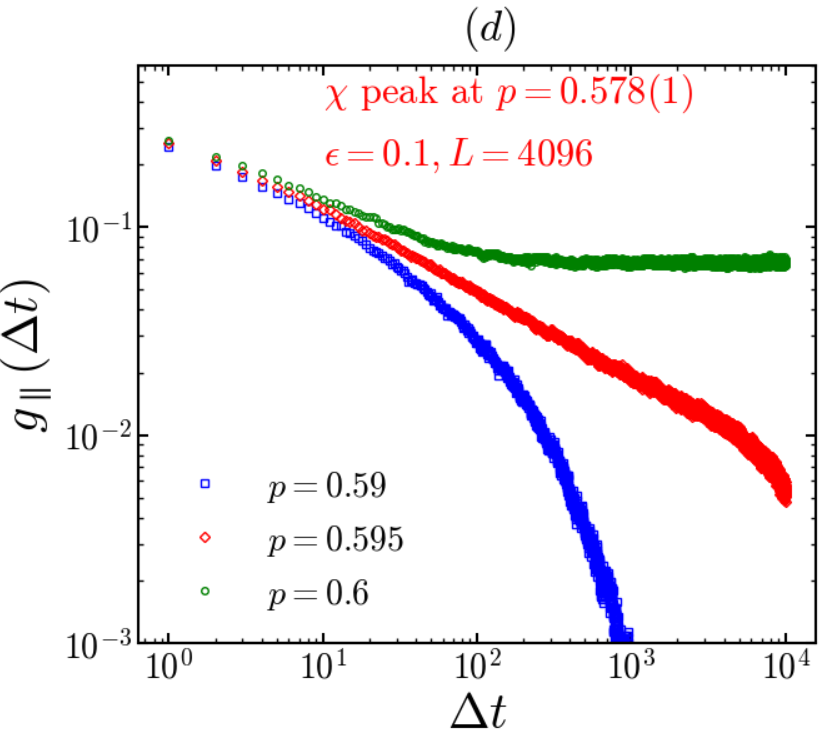}\\
    \caption{ (a) \& (b) Log-log plot of equal time correlation $g_{\perp}(r,t)$ vs $r$ is shown for $\epsilon=0.01$ and $0.1$. (c) \& (d) Log-log plot of autocorrelation $g_{\parallel}(\Delta t)$ vs $\Delta t$ is shown for $\epsilon=0.01$ and $0.1$.  In all the figures, the value of $p$ corresponding to the susceptibility peak is also specified. For a given value of $\epsilon$, the value of $p$ at which power-law nature is observed for the correlations, and dynamic susceptibility $\chi$ shows a maximum, is observed to be slightly different.}
    \label{corr_epsilon}
\end{figure*}

Spatial and temporal correlations are evaluated for nonzero $\epsilon$, and we find that the correlations exhibit power law decay at some specific value of $p$. In Figures.~\ref{corr_epsilon}~(a) and (b), we show the equal time correlation plots $g_\perp(r,t)$ vs $r$ for different $p$ near the pseudo critical point for two different values of $\epsilon$. From Figure.~\ref{corr_epsilon}~(a), we can infer that $g_\perp(r,t)$ nearly follows a straight line at a specific value of $p = 0.611(4)$, indicating power-law decay. Note that the value of $p$ at which this happens is slightly lower than the values of $p$ at which the maximum of $\chi$ is obtained, which is $p = 0.617(1)$. This difference becomes more pronounced at higher values of $\epsilon$ as seen in Figure.~\ref{corr_epsilon}~(b) for $\epsilon = 0.1$.  Similar observations regarding the behaviour of the autocorrelation function $g_\parallel(\Delta t)$ can be made from Figure.~\ref{corr_epsilon}~(c) and (d).

The behavior of correlations seen in Figure~\ref{corr_epsilon} as the control parameter $p$ is varied, is like that of correlations seen in DP and Ising-like models, i.e, a power-law with an exponential cutoff $g(r) \sim r^{-x}e^{-r/\xi}$ where $\xi$ is the correlation length, $x$ is a suitable exponent, and $r$ denotes spatial distance (or $\Delta t$ in the case of autocorrelation). Approaching the pseudo threshold from below, the correlations are well fitted by this form well, with a systematically increasing correlation length $\xi$, indicating that at the threshold, the correlations follow power-laws. Goodness-of-fit tests also prefer a power-law decay over other plausible forms like a stretched exponential at the threshold.

\begin{figure*}
    \centering
    \includegraphics[width=0.49\linewidth]{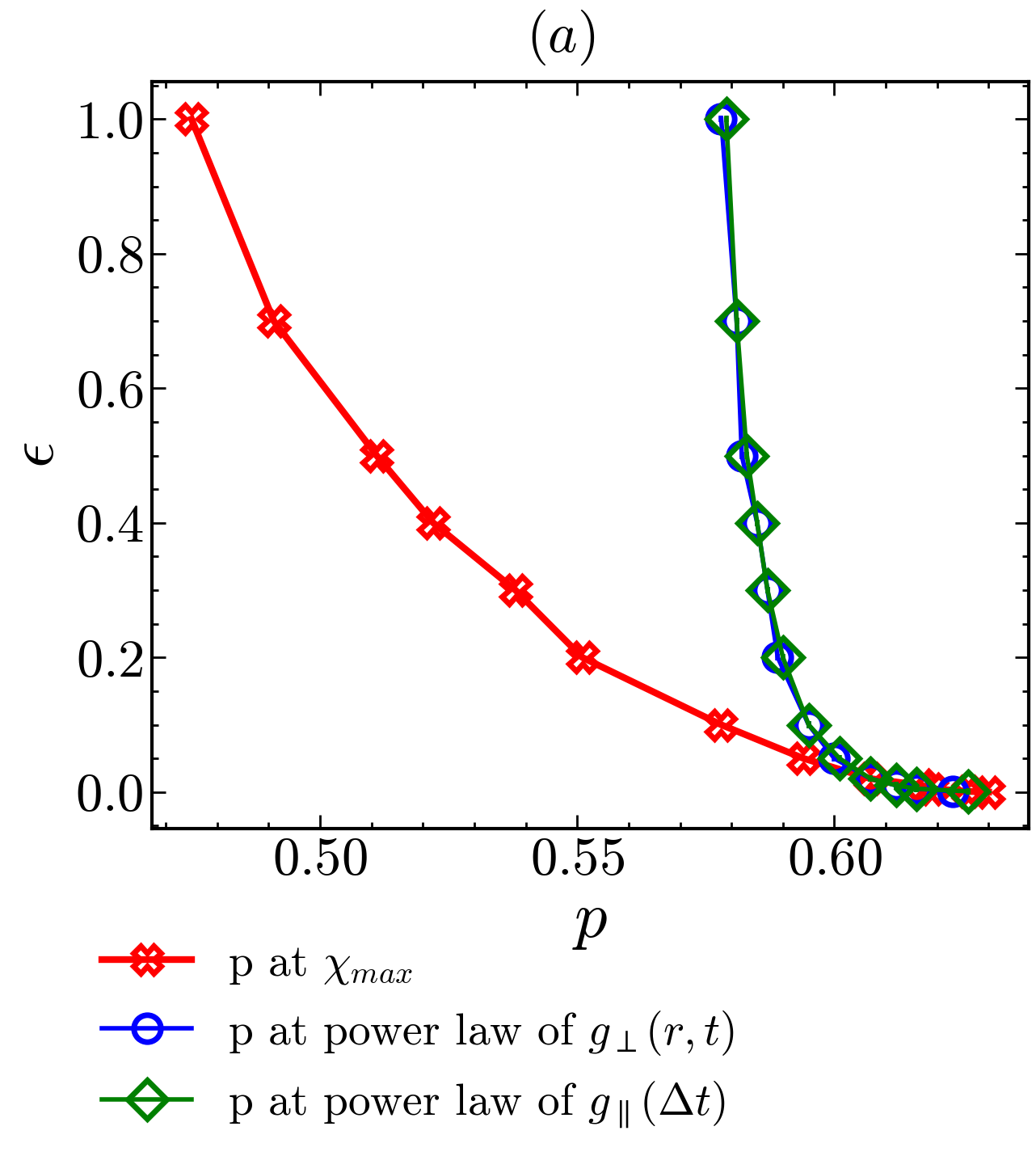}
    \includegraphics[width=0.47\linewidth]{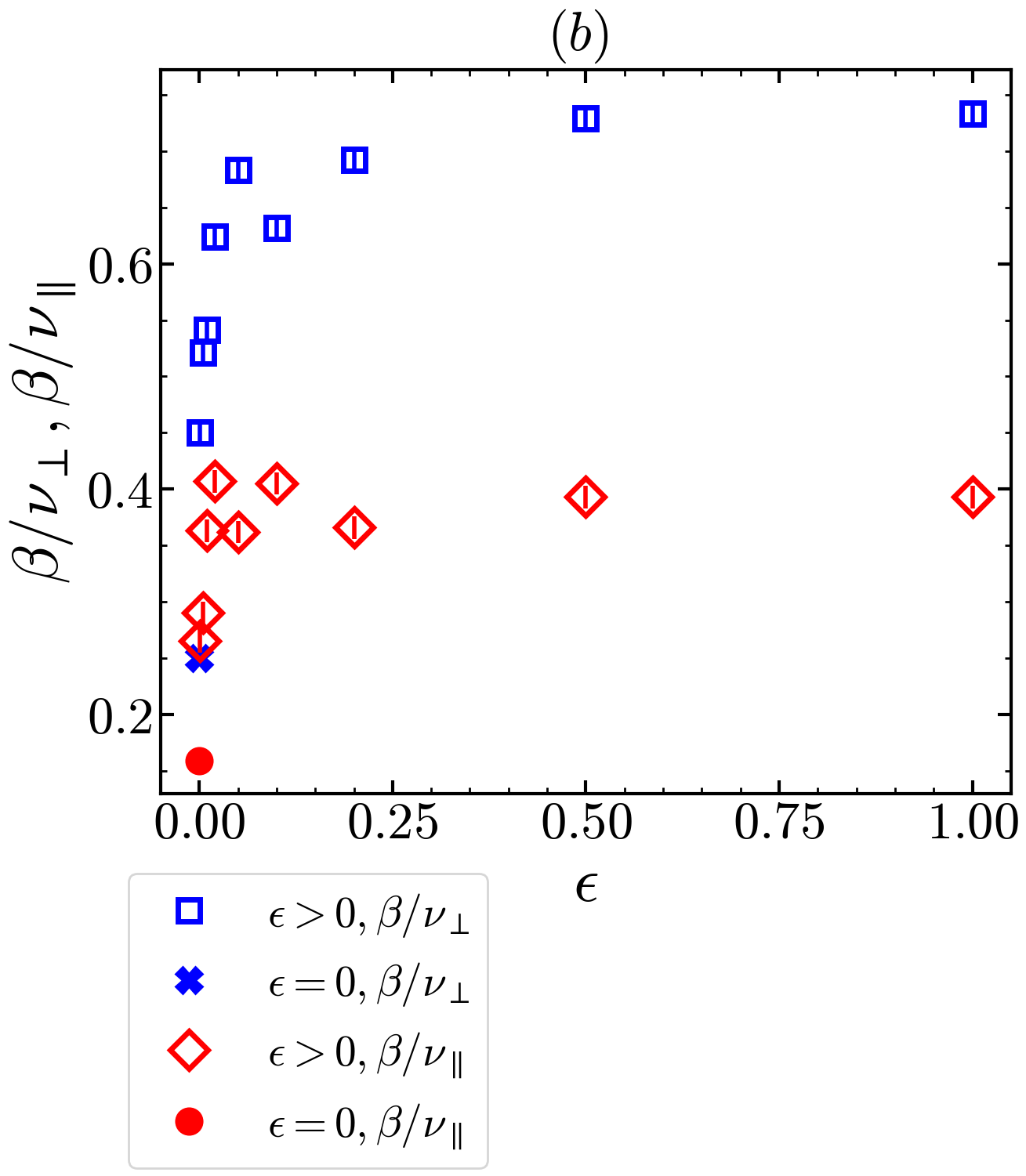}
    \caption{(a) Non-equilibrium Widom line obtained from $p$ corresponding to maxima of dynamical susceptibility $\chi$ and $p$ corresponding to power law behaviour of equal time correlation $g_{\perp}(r,t)$ and auto correlation $g_{\parallel}(\Delta t)$. The two curves coincide only at low $\epsilon$. (b) Variation of the slope of the equal time correlation (exponent $\beta/\nu_{\perp}$) with $\epsilon$ and variation of the slope of the auto correlation (exponent $\beta/\nu_{\parallel}$) with $\epsilon$.  Each point represents the exponent ratio $\beta/\nu_{\perp}$ (blue) or $\beta/\nu_{\parallel}$ (red). The point corresponding to $\epsilon=0$ is marked separately. The slopes are obtained from correlation curves slightly above criticality by fitting a straight line in the initial region ($r\sim (1-500), \Delta t \sim (1-500)$).}
    \label{slope}
\end{figure*}

In Figure.~\ref{slope}~(a), we plot the value of $p$ at which power law decay in spatial and temporal correlations are seen along with the value at which the maximum of dynamic susceptibility is obtained for a given $\epsilon$. We can see that, for any $\epsilon$, the power law decays in spatial and temporal correlations develop at the same value of $p$. For low $\epsilon$, this $p$ value is also where the maximum of the dynamic susceptibility is obtained. However, for higher $\epsilon$, the values of $p$ at which correlations show power law nature and dynamic susceptibility shows a maximum are different as can be seen from Figure.~\ref{slope}~(a). This indicates that spatial and temporal correlations exhibit power-law decays at a value of the control parameter different from the pseudo-threshold corresponding to the peak of the dynamic susceptibility. Thus in the presence of spontaneous activity, there seem to be two  pseudo-thresholds, one where the response function is maximum and another where the spatial and temporal correlations show scale free behaviour. We note that although power-laws are usually associated with the notion of criticality, there are instances where they can arise in non-critical situations as well~\cite{touboul2017,Schulman2021,Faqeeh2019,Priesemann2018}.

Finally, the slope of the power laws in Figure.~\ref{corr_epsilon} gives the exponent ratios $\beta/\nu_\perp$ and $\beta/\nu_\parallel$ whose variations with $\epsilon$ are plotted in Figure.~\ref{slope}~(b) along with that for $\epsilon = 0$ for comparison. We can see that both the exponent ratios show an increasing trend with $\epsilon$ and go to roughly a constant value for larger $\epsilon$ values. 

\section{Conclusions}
\label{section3}
In this work, a three-species dynamical model, which is equivalent to the semi-directed percolation (SDP) problem, is proposed. We study the active-absorbing phase transition in the dynamical model in detail using simulation methods, obtaining the threshold and the critical exponents. We verify earlier theoretical results in Ref.~\cite{HOMartin1985,Knezevic2016} regarding the percolation threshold of SDP problem. The determined values of various critical exponents suggest that the dynamical semi-directed model belongs to the universality class of fully directed percolation.

We study the effect of including spontaneous activity in the dynamic SDP model. Results show that the presence of spontaneous activity gives rise to a finite peak in the dynamic susceptibility at a specific value of the control parameter, indicating the presence of a pseudo threshold. The susceptibility peak broadens as we increase the strength of the spontaneous activity, indicating that, rather than a sharp critical point, the system transitions through a
broader critical-like region. Interestingly, the two pseudo-thresholds, identified from the susceptibility peak and from the power-law nature of equal-time and temporal correlations, seem to be different at high levels of spontaneous activity.

\begin{figure*}[!ht]
    \centering
    \includegraphics[width=0.6\linewidth]{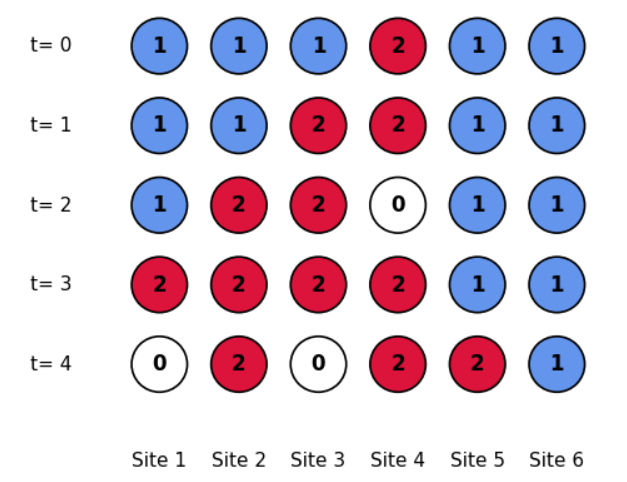}
     \caption{A different dynamic process on a 1D lattice which will generate SDP clusters in time. Here, a currently active site (state 2) can spread activity only if it survives to the next time step, and a site that loses activity remains immune (state 0) for exactly one time step. In a given time step, each active site attempts to spread infection to successive susceptible sites (state 1) on either side with a probability $p$, stopping at the first unsuccessful attempt.}
     \label{schematic}
\end{figure*}

There are alternative ways of realizing the same dynamics; however, they also require three states. For example, in Fig.~\ref{schematic}, we show an SDP cluster generated via a different but equivalent dynamic process to the one defined in Sec.~\ref{section1}. 

There are similarities and differences between the model defined in the paper and the DP-inspired neuronal models \cite{Korchinski2021, WilliamsGarcia2014}. Both are, in fact, three-species models: The refractory period in the case of neuronal models essentially corresponds to an immune state (the site cannot be activated again during the refractory period). 
The difference lies in the fact that, in our problem, active sites remain active or become immune stochastically (Eq.~\ref{a4}). This means that the time for which activity or immunity (refractory period) sustains for a site is probabilistic. Another difference is that, in our problem, a susceptible site can directly become immune (Eq.~\ref{a3}), a feature usually not found in neuronal models (this is like going directly from a susceptible to a refractory state).

In summary, the proposed dynamic model for the semi-directed percolation problem, together with its critical and quasicritical behavior, provides a clearer understanding of the precise similarities and distinctions between isotropic and directed percolation. An open question remains as to whether isotropic critical behavior can be recovered from semi-directed models incorporating spontaneous activity with a distribution other than the uniform case examined here. The observed differences between the pseudo-thresholds call for further investigation. More broadly, embedding a model within a new dynamic framework often enables the application of alternative analytical approaches, yielding fresh insights. We hope that the present study will stimulate further research in this direction.

\ack
\noindent Authors acknowledge the use of the high-performance computing cluster established at Cochin University of Science and Technology (CUSAT) under the Rashtriya Uchchatar Shiksha Abhiyan (RUSA 2.0) scheme (No. CUSAT/PL(UGC).A1/2314/2023, No: T3A).

\section*{Code availability}
Sample source code for our project is available on the GitHub repository.\\
\url{https://github.com/jasnaCK/Semidirected-3species-model}

\end{document}